\documentclass[hyper]{JHEP3} 

\usepackage{epsfig}
\usepackage{amsmath}
\usepackage{cite}

\newcommand{\vect}[1]{\overrightarrow{#1}}
\newcommand{\smbox}[1]{\mbox{\scriptsize #1}}
\newcommand{\tbox}[1]{\mbox{\tiny #1}}

\newcommand{\gL}{g_{\mbox{\tiny{L}}}}
\newcommand{\gR}{g_{\mbox{\tiny{R}}}}
\newcommand{\gY}{g_{\mbox{\tiny{Y}}}}
\newcommand{\gBL}{g_{\mbox{\tiny{BL}}}}
\newcommand{\gdenomA}{\sqrt{\gL^2\gR^2+\gBL^2\gL^2+\gBL^2\gR^2}}
\newcommand{\gdenomB}{\sqrt{\gBL^2+\gR^2}}
\newcommand{\Psibar}{\overline{\Psi}}
\newcommand{\Bvec}{B^{\tbox{(11)}}_{\mu}}
\newcommand{\Bsca}{B^{\tbox{(11)}}_{\tbox{(-)}}}
\newcommand{\beq}{\begin{equation}}
\newcommand{\eeq}{\end{equation}}

\title{\Large\bf Mixed Dark Matter in Universal Extra Dimension Models
with TeV Scale $W_{\tbox{R}}$ and $Z'$}

\author{\bf Ken Hsieh$^{1}$, R.N. Mohapatra$^{1}$ and Salah Nasri$^{2}$\\
$^1$Department of Physics, University of Maryland,
College Park, MD 20742, USA\\
$^2$Department of Physics, University of Florida, Gainesville, FL, 32611\\
E-mail: \email {kenhsieh@physics.umd.edu, rmohapat@physics.umd.edu, snasri@phy.ufl.edu}}

\received{\today}               
\accepted{\today}               

\preprint{UFIFT-HEP-06-16 \\
UMD-HEP-06-055\\ \today}

\abstract{We show that in a class of universal extra dimension (UED)
models that solves both the neutrino mass and proton decay problems
using low scale left-right symmetry, the dark matter of the Universe
consists of an admixture of KK photon and KK right-handed neutrinos.
We present a full calculation of the dark matter density in these
models taking into account the co-annihilation effects due to near
by states such as the scalar partner of the KK photon as well as
fermion states near the right-handed KK neutrino. Using the value of
the relic CDM density, we obtain upper limits on $R^{-1}$ of about
$400-650$ GeV and $M_{Z'}\leq 1.5$ TeV, both being accessible to
LHC.  For a region in this parameter space where the KK right-handed
neutrino contributes significantly to the total relic density of
dark matter, we obtain a lower bound on the dark matter-nucleon
scattering cross section of $10^{-44}$ cm$^2$, which can be probed
by the next round of dark matter search experiments.}


\begin{document}

\section{Introduction}
\label{sec:intro}

Understanding the dark constituent of the Universe is one of the
major problems of physics beyond the Standard Model (SM). While in the
supersymmetric extensions of the Standard Model, the lightest
supersymmetric partner (LSP) of the standard model fields is one of the
most well motivated candidates for the cold dark matter (CDM), it is
by no means unique and other viable CDM candidates have been
proposed in the literature \cite{tait,aghase,mdm}. It is hoped
that the Large Hadron Collider (LHC) will provide evidence for
supersymmetry making the case for this particle stronger.
Nonetheless, at this point, different candidates must be studied in
order to isolate their possibly different signatures in other
experiments in order to make a proper identification of the true
candidate. With this goal in mind, in this paper, we continue our
study \cite{hsieh} of a class of dark matter
candidates \cite{tait,kong}, which arises in models with extra
dimensions\cite{Antoniadis:1990ew,antoniadis,arkani-hamed}, the so-called universal
extra dimensional (UED) models \cite{acd}.

The UED models lead to a very different kind of TeV scale physics
and will also be explored at LHC.  These models have hidden extra
spatial dimensions with sizes of order of an inverse TeV with all
SM fields residing in all the dimensions. There could be
one or two such extra dimensions and they are compactified with
radius $R^{-1}\leq $ TeV \cite{acd}. It has recently been pointed out
\cite{tait} that the lightest Kaluza-Klein (KK) particles of these models
being stable can serve as viable dark matter candidates. This result
is nontrivial due to the fact that the dark matter relic abundance
is determined by the interactions in the theory which are
predetermined by the Standard Model. It turns out that in the minimal, 5D extra
dimension UED models based on the standard model gauge group, the first
KK mode of the hypercharge boson is the
dark matter candidate provided the inverse size of the extra
dimension is less than a TeV \cite{tait}.

A generic phenomenological problem with 5D UED models based on the
Standard Model gauge group is that they can lead to rapid proton
decay as well as unsuppressed neutrino masses.  One way to cure the
rapid proton decay problem is to consider six dimensions \cite{yee}
where the two extra spatial dimensions lead to a new $U(1)$ global
symmetry that suppresses the strength of all baryon number
nonconserving operators. On the other hand both the neutrino mass
and the proton decay problem can be solved simultaneously if we
extend the gauge group of the six dimensional model to
$SU(2)_L\times SU(2)_{\tbox{R}}\times U(1)_{B- L}$ \cite{abdel}. This avoids
having to invoke a seventh warped extra dimension solely for the
purpose of solving the neutrino mass problem \cite{app}. With
appropriate orbifolding, a neutrino mass comes out to be of the
desired order due to a combination two factors: the existence of
$B-L$ gauge symmetry and the orbifolding that keeps the left-handed
singlet neutrino as a zero-mode which forbids the lower dimensional
operators that could give unsuppressed neutrino mass.  Another advantage
of the 6D
models over the 5D ones is that cancellation of gravitational anomaly
automatically leads to the existence of the right-handed
neutrinos \cite{poppitz} needed for generating neutrino masses.

In a recent paper \cite{hsieh}, we pointed out that the 6D UED models
with an extended gauge group \cite{abdel} provide a
two-component picture of dark matter consisting of a KK right-handed
neutrino and a KK hypercharge boson. We presented a detailed
calculation of the relic abundance of both the
$\nu^{\tbox{KK}}_{\smbox{R}}$ and the
$B_Y^{\tbox{KK}}$ as well as the cross section for scattering of the dark
matter in the cryogenic detectors in these models. The two main
results of this calculation\cite{hsieh} are that: (i) present
experimental limits on the value of
the relic density \cite{wmap} imply very stringent limits on the the
two fundamental parameters of the theory i.e. $R^{-1}$ and the
second $Z'$-boson associated with the extended gauge group i.e.
$R^{-1}\leq 550$ GeV and $M_{Z'}\leq 1.2$ TeV and (ii) for one
particular region in
this parameter range where the relic density of the KK right-handed neutrino
contributes significantly to the total relic density of the dark matter,
the DM-nucleon cross-section is greater than $10^{-44}$
cm$^2$, and is accessible to the next round of dark
matter searches.  Thus combined with LHC results for an extra
$Z'$ search, the direct dark matter search experiments could
rule out this model. This result is to be contrasted
with that of minimal 5-D UED models, where the above experiments
will only rule out a part of the parameter space. Discovery of two
components
to dark matter should also have implications for cosmology of structure
formation.

In this paper, we extend the work of ref.\ \cite{hsieh} in several
ways: (i) we update our calculations taking into account the
co-annihilation effect of nearby states; (ii) a feature unique to
six and higher dimensional models is the presence of physical scalar
KK states of gauge bosons degenerate at the tree level with
$\gamma_{\tbox{KK}}$ state and will therefore impact the discussion of KK
dark matter.  Its couplings to matter have different Lorentz
structure and therefore contribute in different ways to the relic
density.  We discuss the relative significance of the scalar state
and its effect on the relic density calculation of the previous
paper \cite{hsieh} for both the cases when it is lighter and heavier
than the $\gamma_{\tbox{KK}}$ state. (iii) We also comment on the extra $W$
and $Z'$ boson phenomenology in the model.

This paper is organized as follows: in Section \ref{sec:setup}, we
review the basic
set up of the model \cite{abdel}.  In Section \ref{sec:spectrum}, we
present the
 spectrum of states at tree level.  In Sections \ref{sec:DM1} and
\ref{sec:DM2}, we discuss the
 relic density of $\nu_{KK}$ states and the hypercharge vector and
pseudoscalar, respectively.
 In Section \ref{sec:RelicResult}, we give the overall picture of dark
matter in these models in
terms of relic abundance and rates of direct detection.
 Section \ref{sec:DirectD} discusses the signals such two-component dark
matter would give
in direct detection experiments.
In Section \ref{sec:WZPheno}, we give the phenomenology of the model for
colliders,
especially the $Z'$ and $W_{\tbox{R}}$ production and decays.
Finally, in Section \ref{sec:Conclusions} we
present our conclusions.
\section{Set up of the Model}
\label{sec:setup}
We choose the gauge group of the model to be $SU(3)_c\times
SU(2)_L\times SU(2)_{\tbox{R}}\times U(1)_{\tbox{B-L}}$ with matter content per
generation as follows:
\begin{eqnarray}
{\cal Q}_{1,-}, {\cal Q}'_{1,-}= (3,2,1,\tfrac{1}{3});&
{\cal Q}_{2,+}, {\cal Q}'_{2,+}= (3,1,2,\tfrac{1}{3});\nonumber\\
{\cal \psi}_{1,-}, {\cal \psi}'_{1,-}= (1,2,1,-1);& {\cal
\psi}_{2,+}, {\cal \psi}'_{2,+}= (1,1,2,-1); \label{matter}
\end{eqnarray}
where, within parenthesis, we have written the quantum numbers that
correspond to each group factor, respectively and the subscript
gives the six dimensional chirality to cancel gravitational anomaly
in six dimensions. We denote the gauge bosons as $G_M$,
$W^{\pm}_{1,M}$, $W^{\pm}_{2,M}$, and $B_M$, for $SU(3)_c$,
$SU(2)_L$, $SU(2)_{\tbox{R}}$ and $U(1)_{\tbox{B-L}}$ respectively, where
$M=0,1,2,3,4,5$ denotes the six space-time indices. We will also use
the following short hand notations:  Greek letters
$\mu,\nu,\dots=0,1,2,3$ to denote  usual four dimensions indices, as
usual, and lower case Latin letters $a,b,\dots=4,5$ for those of the
extra space dimensions. We will also use $\vec y$ to denote the
($x_4,x_5$) coordinates of a point in the extra space.

First, we compactify the extra $x_4$, $x_5$ dimensions into
a torus, $T^2$, with equal radii, $R$, by imposing  periodicity
conditions, $\varphi(x_4,x_5) = \varphi(x_4+ 2\pi R,x_5) =
\varphi(x_4,x_5+ 2\pi R)$ for any field $\varphi$.
 This has the effect of breaking the original $SO(1,5)$
Lorentz symmetry group of the six dimensional space  into the
subgroup $SO(1,3)\times Z_4$, where the last factor corresponds to
the group of discrete  rotations in the $x_4$-$x_5$ plane, by angles
of $k\pi/2$ for $k=0,1,2,3$. This is a subgroup of the continuous
$U(1)_{45}$ rotational symmetry contained in $SO(1,5)$. The
remaining $SO(1,3)$ symmetry  gives the usual 4D Lorentz invariance.
The presence of the surviving $Z_4$ symmetry leads to suppression of
proton decay \cite{yee} as well as neutrino mass \cite{abdel}.
\newline
\indent Employing the further orbifolding conditions :
\begin{eqnarray}
Z_2&:& {\bf{y}} \rightarrow -{\bf {y}}
 \\ \nonumber
Z'_2&:& \left\{ \begin{array}{ll}
        (x_4,x_5)~' \rightarrow - (x_4,x_5)~'  \\
        {\bf{y}}~' = {\bf{y}} - (\pi R /2, \pi R/2)&
        \end{array} \right.
\end{eqnarray}
We can project out the
zero modes and obtain the KK modes by assigning appropriate
$Z_2\times Z'_2$ quantum numbers to the fields.

In the effective 4D theory the mass of each mode has the
form: $m_{N}^2 = m_0^2 + \frac{N}{R^2}$; with $N=\vec{n}^2=n_1^2 +
n_2^2$ and $m_0$ is the Higgs vacuum expectation value (vev)
contribution to mass, and the physical mass of the zero mode.

We assign the following $Z_2\times Z'_2$ charges to the
various fields:
\begin{eqnarray}
G_\mu(+,+);\ B_\mu(+,+);\ W_{1,\mu}^{3,\pm}(+,+);\
W^3_{2,\mu}(+,+);\ W^\pm_{2,\mu}(+,-); \nonumber\\
G_{a}(-,-);\ B_a(-,-);\ W_{1,a}^{3,\pm}(-,-);\ W^3_{2,a}(-,-);\
W^\pm_{2,a}(-,+). \label{gparity}
\end{eqnarray}
For quarks we choose,
\begin{eqnarray}
Q_{\tbox{1L}}\!\equiv\!
\left(\begin{array}{c} u_{\tbox{1L}}(+,+)\\
d_{\tbox{1L}}(+,+)\end{array}\right);
 \
Q'_{\tbox{1L}}\!\equiv\!
\left(\begin{array}{c} u'_{\tbox{1L}}(+,-)\\
d'_{\tbox{1L}}(+,-)\end{array}\right);
 \
Q_{\tbox{1R}}\!\equiv\!
\left(\begin{array}{c} u_{\tbox{1R}}(-,-)\\
d_{\tbox{1R}}(-,-)\end{array}\right);
\
Q'_{\tbox{1R}}\!\equiv\!
\left(\begin{array}{c} u'_{\tbox{1R}}(-,+)\\
d'_{\tbox{1R}}(-,+)\end{array}\right);
\nonumber\\
Q_{\tbox{2L}}\!\equiv\!
\left(\begin{array}{c} u_{\tbox{2L}}(-,-)\\
d_{\tbox{2L}}(-,+)\end{array}\right);
 \
Q'_{\tbox{2L}}\!\equiv\!
\left(\begin{array}{c} u'_{\tbox{2L}}(-,+)\\
d'_{\tbox{2L}}(-,-)\end{array}\right);
 \
Q_{\tbox{2R}}\!\equiv\!
\left(\begin{array}{c} u_{\tbox{2R}}(+,+)\\
d_{\tbox{2R}}(+,-)\end{array}\right);
\
Q'_{\tbox{2R}}\!\equiv\!
\left(\begin{array}{c} u'_{\tbox{2R}}(+,-)\\
d'_{\tbox{2R}}(+,+)\end{array}\right);
\label{eq:quarks}
\end{eqnarray}
and for leptons:
\begin{eqnarray}
\psi_{\tbox{1L}}&\!\equiv\!
\left(\begin{array}{c} \nu_{\tbox{1L}}(+,+)  \\
e_{\tbox{1L}}(+,+)\end{array}\right);
 \ \
\psi'_{\tbox{1L}}\!\equiv\!
\left(\begin{array}{c} \nu'_{\tbox{1L}}(-,+)  \\
e'_{\tbox{1L}}(-,+)\end{array}\right);
 \ \
\psi_{\tbox{1R}}\!\equiv\!
\left(\begin{array}{c} \nu_{\tbox{1R}}(-,-)  \\
e_{\tbox{1R}}(-,-)\end{array}\right);
 \ \
\psi'_{\tbox{1R}}\!\equiv\!
\left(\begin{array}{c} \nu'_{\tbox{1R}}(+,-) \\
e'_{\tbox{1R}}(+,-)\end{array}\right);
\nonumber \\
\psi_{\tbox{2L}}&\!\equiv\!
\left(\begin{array}{c} \nu_{\tbox{2L}}(-,+) \\
e_{\tbox{2L}}(-,-)\end{array}\right);
 \ \
\psi'_{\tbox{2L}}\!\equiv\!
\left(\begin{array}{c} \nu'_{\tbox{2L}}(+,+)\\
e'_{\tbox{2L}}(+,-)\end{array}\right);
 \ \
\psi_{\tbox{2R}}\!\equiv\!
\left(\begin{array}{c} \nu_{\tbox{2R}}(+,-)\\
e_{\tbox{2R}}(+,+)\end{array}\right);
 \ \
\psi'_{\tbox{2R}}\!\equiv\!
\left(\begin{array}{c} \nu'_{\tbox{2R}}(-,-)\\
e'_{\tbox{2R}}(-,+)\end{array}\right).
\label{eq:leptons}
\end{eqnarray}
The zero modes i.e. (+,+) fields corresponds to the standard model
fields along with an extra singlet neutrino which is left-handed.
They will have zero mass prior to gauge symmetry breaking. The
singlet neutrino state being a left-handed (instead of right-handed
as in the usual case) has important implications for neutrino mass.
For example, the conventional Dirac mass term $\bar{L}H\nu_{\tbox{R}}$ is not
present due to the selection rules of the model and Lorentz
invariance. Similarly, $L\tilde{H}\nu_{2L}$ is forbidden by gauge
invariance as is the operator $(LH)^2$. Thus neutrino mass comes only
from much higher dimensional terms.

%

For the Higgs bosons, we choose a bidoublet, which will
be needed to give masses to fermions and break the standard model
symmetry and and a pair of doublets $\chi_{L,R}$ with the following
$Z_2\times Z'_2$ quantum numbers:
\begin{align}
\phi\equiv
\left(\begin{array}{cc} \phi^0_u(+,+) & \phi^+_d(+,-)\\
\phi^-_u(+,+) &  \phi^0_d(+,-)\end{array}\right);\quad
\chi_L\equiv \left(\begin{array}{c} \chi^0_L(-,+) \\
\chi^-_L(-,+)\end{array}\right);\quad
\chi_{\tbox{R}}\equiv \left(\begin{array}{c} \chi^0_{\tbox{R}}(+,+) \\
\chi^-_{\tbox{R}}(+,-)\end{array}\right),
\end{align}
and the following charge assignment under the gauge group,
\begin{eqnarray}
\phi &=& (1,2,2,0),\nonumber\\
\chi_L&=&(1,2,1,-1),\quad \chi_{\tbox{R}}=(1,1,2,-1).
\end{eqnarray}
At the zero mode level, only the SM doublet $(\phi^0_u, \phi^-_u)$
 and a singlet $\chi^0_{\tbox{R}}$ appear.  The vacuum expectation values
(vev) of these fields, namely
 $\langle\phi^0_u\rangle = v_{wk}$ and $\langle\chi^0_{\tbox{R}}\rangle=
v_{\tbox{R}}$,
break the SM symmetry and the extra $U(1)_Y'$ gauge group,
respectively.
A diagram that illustrates the lowest KK modes of all the particles
and their masses is shown in Fig.~\ref{fig:level} with the following
identification of modes in Table \ref{modetable}.

\TABLE{
\begin{tabular}{|c|l|}
\hline ($Z_2,Z_2^{\prime}$) & Particle Content \\
\hline (++) &
$\begin{array}{l}
Q_{\tbox{1L}};~
u_{\tbox{2R}};~
d^{\prime}_{\tbox{2R}};~
\psi_{\tbox{1L}};~
e_{\tbox{2R}};~
\nu^{\prime}_{\tbox{2L}};
\vspace{0.05in}
\\
G_{\mu};~
B_{\mu};~
W_{1,\mu}^{3,\pm};~
W_{2,\mu}^3;
\vspace{0.05in}
\\
\phi_u^0;~
\phi_u^-;~
\chi_{\tbox{R}}^{\tbox{0}}
\vspace{0.05in}
\end{array}$
\\
\hline ($+-$) &
$\begin{array}{l}
Q^{\prime}_{\tbox{1L}};~
u^{\prime}_{\tbox{2R}};~
d_{\tbox{2R}};~
\psi^{\prime}_{\tbox{1R}};~
\nu_{\tbox{2R}};~
e^{\prime}_{\tbox{2R}};
\vspace{0.05in}
\\
W_{2,\mu}^{\pm};
\vspace{0.05in}
\\
\phi_d^+;~
\phi_d^{\tbox{0}};~
\chi_{\tbox{R}}^{\tbox{-}}
\vspace{0.05in}
\end{array}$
\\
\hline ($-+$) &
$\begin{array}{l}
Q^{\prime}_{\tbox{1R}};~
u^{\prime}_{\tbox{2L}};~
d_{\tbox{2L}};~
\psi^{\prime}_{\tbox{1L}};~
\nu_{\tbox{2L}};~
e^{\prime}_{\tbox{2R}};
\vspace{0.05in}
\\
W_{2,a}^{\pm};
\vspace{0.05in}
\\
\chi_{\tbox{L}}^{\tbox{0}};~
\chi_{\tbox{L}}^{\tbox{-}}
\vspace{0.05in}
\end{array}$
\\
\hline
($--$)
&
$\begin{array}{l}
Q_{\tbox{1R}};~
u_{\tbox{2L}};~
d^{\prime}_{\tbox{2L}};~
\psi_{\tbox{1R}};~
\nu_{\tbox{1R}};~
e_{\tbox{2L}};
\vspace{0.05in}
\\
G_a;~
B_a;~
W_{1,a}^{3,\pm};~
W_{2,a}^{3}
\vspace{0.05in}
\end{array}$
\\
\hline
\end{tabular}
 \caption{Particle content of 6D model separated by $Z_2\times
Z_2^{\prime}$ parities.}
\label{modetable}
}

\FIGURE[h]{
\includegraphics[width=5in]{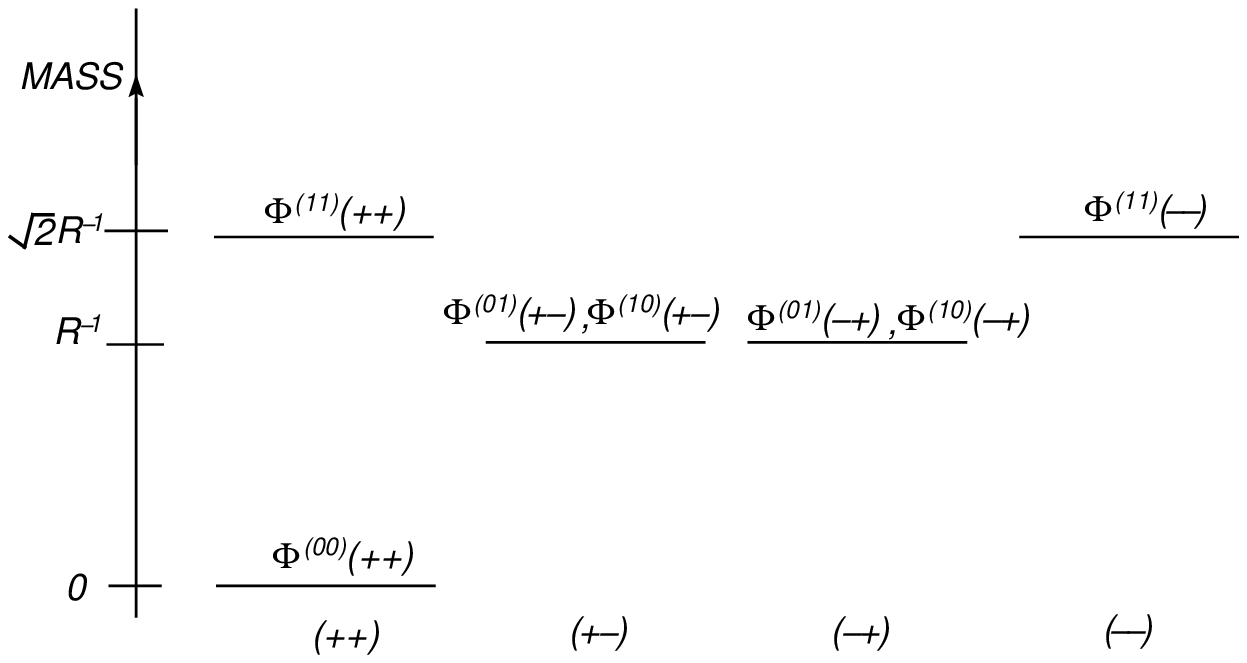}
\caption{The masses of lowest KK-modes of 6D model.}
\label{fig:level}}
%

The most general Yukawa couplings in the model are
\beq
h_u \bar Q_1\phi Q_2 + h_d \bar Q_1\tilde\phi Q'_2 +
h_e\bar \psi_1\tilde \phi \psi_2 +
h'_u \bar Q'_1\phi Q'_2 + h'_d \bar Q'_1\tilde\phi Q_2 +
h'_e\bar \psi'_1\tilde \phi \psi'_2 +
 h.c.;
\label{Yuk}
\eeq
where $\tilde \phi\equiv \tau_2\phi^* \tau_2$ is
the charge conjugate field of $\phi$. A six dimensional realization
of the left-right symmetry, which interchanges  the subscripts:
$1\leftrightarrow 2$, is obtained provided the $3\times 3$ Yukawa
coupling matrices satisfy the constraints: $h_u~=~h^{\dagger}_u$;
$h'_u~=~h^{'\dagger}_u$; $h_e~=~h^{\dagger}_e;$
$h'_e~=~h^{'\dagger}_e;$ $h_d~=~h^{'\dagger}_d$. At the zero mode
level one obtains the SM Yukawa couplings
\beq {\cal L}~=~h_u \bar
Q\phi_u u_R + h_d \bar Q\tilde\phi_u d_R + h_e \bar L\tilde \phi_u
e_R + h.c.
\label{yukawa}
\eeq
It is important to notice that in the
above equation $h_{u,e}$ are hermitian matrices, while $h_d$ is
not. The vev of $\phi_u$ gives mass to the charged fermions of the model.
As far as the neutrino mass is concerned,
the lowest dimensional gauge invariant operator in six-D that gives rise
to neutrino mass after compactificaion has the
form $\psi^T_{1,L}{\phi}\psi_{2,L}\chi^2_{R}$ and leads to neutrino mass
$m_\nu\simeq \lambda \frac{v_{wk}v^2_R}{M^2_*(M^2_*R)^3}$. For $M_*R\sim
100$ and $M_*\sim$ 10 TeV, $v_R\sim 2 $ TeV and $\lambda\sim 10^{-3}$, we
get neutrino masses of order $\sim$ eV without fine tuning. Furthermore,
it predicts that the neutrino mass is Dirac (predominantly) rather than
Majorana type.

As there are a large number of KK modes, one may worry whether or
not electroweak precision constraints in terms of S and T parameters
are satisfied. It has been shown that in the minimal universal extra
dimension (MUED) the KK contributions to the T parameter almost
cancel for heavier standard model Higgs \cite{acd, gm}. However, it
was found that in the MUED for Higgs mass heavier than 300 GeV
the lightest Kaluza-Klein particle is the charged KK Higgs
\cite{KMS}. The abundance of such charged massive particles are
inconsistent with big bang nucleosynthesis as well as other cosmological
observations for masses less than a
TeV \cite{KY}. This lead to the conclusion that the compactification
scale $1/R >$ 400 GeV for $m_H >$ 300 GeV. To our knowledge, there
has been no such analysis for the $6-$D models similar to ours, and
it is outside the scope of the current paper to perform a complete
analysis regarding the electroweak constraints. Therefore, we leave
the investigation of this open issue for future work.

\section{Spectrum of Particles}
\label{sec:spectrum}
Once the extra dimensions are compactified, the KK modes are labelled by the quanta of momenta
in the extra dimensions.  As we have two such extra spatial dimensions, the KK modes are labelled
by two integers, and we will denote a KK mode as $\phi^{(mn)}$, where $m$ ($n$) is the
momentum in the quantized unit of $R^{-1}$ along the fifth (sixth) dimension.  A detailed
expansion of a field in the 6D theory into KK mode is presented in the Appendix.  Generally,
$\phi^{(mn)}$ would receive a (mass)$^2$ of the order $(m^2+n^2)R^{-2}$.

\subsection{Gauge and Higgs Particles at the Zeroth KK Level}
In the gauge basis, we have the zero-mode gauge bosons:
$B^{\smbox{(00)}}_{\tbox{(B-L)}\mu},W^{\pm,3\smbox{(00)}}_{L,\mu}$, and
$W^{3\smbox{(00)}}_{R,\mu}$.  After symmetry-breaking, we will
have the usual SM gauge bosons: one exactly massless gauge boson, $A_{\mu}^{\smbox{(00)}}$, one
pair of massive, charged vector boson $W_{L,\mu}^{\pm,\smbox{(00)}}$, and
one massive neutral guage boson $Z_{\mu}^{\smbox{(00)}}$.
In addition, we will have another neutral gauge boson
$Z_{\mu}^{\prime\smbox{(00)}}$, as well as mixing between $Z_{\mu}^{\smbox{(00)}}$
and $Z_{\mu}^{\prime\smbox{(00)}}$.

In this subsection we calculate the zeroth-mode gauge boson masses and mixings from Higgs
mechanism (and drop the $(00)$ superscript throughout this
subsection). The relevant terms are
\begin{align}
\mathcal{L}_h=\mbox{Tr}[(D_{\mu}\phi)^{\dag}D_{\mu}\phi]+
(D^{\mu}\chi_{\tbox{R}})^{\ast}D_{\mu}\chi_{\tbox{R}}
+(D^{\mu}\chi_L)^{\ast}D_{\mu}\chi_L
\end{align}
where
\begin{align}
D_{\mu}\phi&=\partial_{\mu}\phi-i\gL
(\vect{\tau}\cdot\vect{W_L}_{\mu})\phi+i
\gR\phi(\vect{\tau}\cdot\vect{W_{\tbox{R}}}_{\mu}), \nonumber\\
\phi&=\begin{pmatrix}\phi_u^0 & \phi_d^+  \\ \phi_u^- &
\phi_d^0\end{pmatrix},\nonumber\\
D_{\mu}\chi_L&=\left(\partial_{\mu}-i\gL
(\vect{\tau}\cdot\vect{W_L,}_{\mu}) +i(\frac{1}{2})\gBL
B_{\tbox{(B-L)},\mu}\right)
\begin{pmatrix}\chi_L^0 \\ \chi_L^-\end{pmatrix},\nonumber\\
D_{\mu}\chi_{\tbox{R}}&=\left(\partial_{\mu}-i\gR
(\vect{\tau}\cdot\vect{W_{\tbox{R}},}_{\mu}) +i(\frac{1}{2})\gBL
B_{\tbox{(B-L)},\mu}\right)
\begin{pmatrix}\chi_{\tbox{R}}^0 \\ \chi_{\tbox{R}}^-\end{pmatrix},\nonumber\\
\vect{\tau}\cdot\vect{W_{\mu}}&=\frac{1}{2}
\begin{pmatrix}W^3_{\mu} & \sqrt{2}W_{\mu}^+  \\ \sqrt{2}W_{\mu}^- &
-W^3_{\mu}\end{pmatrix}.
\end{align}
With vev of the fields $\langle\phi_u^0\rangle=v_w$  and
$\langle\chi_{\tbox{R}}^0\rangle=v_{\tbox{R}}$, we obtain the following mass terms for
the gauge bosons:
\begin{align}
\mathcal{L}&=\frac{1}{2}v^2_w(W^+_{L,\mu}W^-_{L,\mu})
+\frac{1}{2}(v^2_w+v^2_{\tbox{R}})(W^+_{R,\mu}W^-_{R,\mu})\nonumber\\
&+\frac{1}{2}\begin{pmatrix}W^3_{L,{\mu}} & W^3_{R,{\mu}} &
B_{\tbox{(B-L)}\mu}
\end{pmatrix}
\begin{pmatrix}\frac{1}{2}\gL^2v_w^2 & -\frac{1}{2}\gL\gR v_w^2 & 0 \\
-\frac{1}{2}\gL\gR v_w^2 & \frac{1}{2}\gR^2(v_w^2+v_{\tbox{R}}^2) &
-\frac{1}{2}
(\gR\gBL)v_{\tbox{R}}^2\\
0& -\frac{1}{2}(\gR\gBL)v_{\tbox{R}}^2& \frac{1}{2}\gBL^2 v_{\tbox{R}}^2\end{pmatrix}
\begin{pmatrix}W^{3,\mu}_{L} \\ W^{3,\mu}_{R} \\ B_{\tbox{(B-L)}}^{\mu} \end{pmatrix}.
\label{eq:HHWW}
\end{align}
The exact expressions of the mass eigenvalues and the compositions
of the eigenstates $(A_{\mu}, Z_{\mu},Z^{\prime}_{\mu})$ in terms of
$(B_{\tbox{(B-L)}\mu},W^3_{L,\mu},W^{3}_{R,\mu})$ are rather complicated,
and we make the approximation of $v_{\tbox{R}} \gg v_w$. In this
approximation, we find the relations,
\begin{align}
\begin{pmatrix} A_{\mu} \\ Z_{\mu} \\ Z^{\prime}_{\mu} \end{pmatrix}
= U^{\dag}_G \begin{pmatrix} W^0_{1,\mu} \\ W^0_{2,\mu} \\ B_{\mu}
\end{pmatrix},
\end{align}
where
\begin{align}
U^{\dag}_G=
\begin{pmatrix}
  \sin\theta_w & \cos\theta_w & 0 \\
  \cos\theta_w & - \sin\theta_w & 0 \\
  0 & 0 & 1
\end{pmatrix}
\begin{pmatrix}
  1 & 0 & 0 \\
  0 & \sin\theta_{\tbox{R}} & \cos\theta_{\tbox{R}} \\
  0 & \cos\theta_{\tbox{R}} & -\sin\theta_{\tbox{R}}
\end{pmatrix},
\end{align}
and
\begin{align}
 \tan\theta_{\tbox{R}}\equiv\frac{\gBL}{\gR}, \quad
\gY^2\equiv\frac{\gBL^2\gR^2}{\gBL^2+\gR^2}, \quad\mbox{and} \quad
\tan\theta_w\equiv \frac{\gY}{\gL}.
\end{align}
It is easy to understand $U_G$ intuitively.  In the limit $v_w\ll
v_{\tbox{R}}$, the symmetry-breaking occurs in two stages, corresponding to
the two matrices in $U_G$. First,  we have $SU(2)_L\times
SU(2)_{\tbox{R}}\times U(1)_{\tbox{B-L}}\rightarrow SU(2)_L\times U(1)_Y$, where a
linear combination of $B_{\tbox{(B-L)},\mu}$ and $W^3_{R,\mu}$ acquire a
mass to become  $Z_{\mu}^{\prime}$, while the orthogonal
combination, $B_{Y,\mu}$, remains massless and serves as the gauge
boson of the residual group $U(1)_Y$.  Then we have the standard
electroweak breaking of $SU(2)_L\times U(1)_Y\rightarrow
U(1)_{\smbox{em}}$, giving us massive $Z_{\mu}$ and the massless
photon $A_{\mu}$.
Using $U_G$, we can simplify the mass matrix enormously
\begin{align}
U^{\dag}_G \mathcal{M}^2 U_G &=
\begin{pmatrix}
0 & 0 & 0 \\
0 & M_Z^2 &
-\frac{\gR^2}{\sqrt{(\gL^2+\gY^2)(\gR^2+\gBL^2)}} M_Z^2 \\
0 & -\frac{\gR^2}{\sqrt{(\gL^2+\gY^2)(\gR^2+\gBL^2)}} M_Z^2 &
M_{Z^{\prime}}^2
\end{pmatrix},
\label{eq:gaugematrix}
\end{align}
where we have defined the mass eigenvalues
(up to $\mathcal{O}(v_w/v_{\tbox{R}})^2$)
\begin{align}
M_Z^2&=\left( \gL^2+\gY^2\right)\frac{v^2_w}{2}\nonumber\\
M_{Z^{\prime}}^2&=(\gBL^2+\gR^2)\frac{v^2_{\tbox{R}}}{2}+
\frac{\gR^4}{(\gBL^2+\gR^2)}\frac{v_w^2}{2}. \label{eq:gaugeeigen}
\end{align}
Here we see that we have explicitly decoupled $A_{\mu}$, and it
remains massless exactly.  Although we have defined $M_Z$ to be same
as the tree-level mass of $Z$-boson of the Standard Model, here
$Z_{\mu}$ is strictly speaking not an eigenstate because of the
$Z-Z^{\prime}$ mixing.  Such mixing would be important, as we will see,
for the calculation of relic density and the direct detection rates of
the dark matter of the model.  However, in the limit $v^2_{\tbox{R}}\gg v^2_w$ that we will be working
with, we can treat the defined masses and states in Eq.~\ref{eq:gaugeeigen} as
eigenvalues and eigenstates,
and treat the mixing terms perturbatively in powers of $(v_w^2/v_{\tbox{R}}^2)$.


The only zero-mode Higgs bosons in the model  are
$\phi_u^{0\smbox{(00)}}, \phi_u^{-\smbox{(00)}},$ and
$\chi_{\tbox{R}}^{0,\smbox{(00)}}$.  Four of the six degrees of freedoms are
eaten and the remaining physical Higgs particles are the real parts
of $\phi_u^{0\smbox{(00)}}$  and $\chi_{\tbox{R}}^{0,\smbox{(00)}}$. The
masses are these particles are determined from the potential, and
are free parameters, whose values, however, do not affect the calculations
of the relic density and direct detection rates of the dark matter.

\subsection{Gauge and Higgs Particles at the First KK Level}
We first consider the question of whether KK modes of Higgs bosons
acquire vevs. The zero modes Higgs bosons acquire vevs due to
negative mass-squared terms in the potential. The higher KK modes of
the Higgs bosons $\phi^{\smbox{(mn)}}$, however, have an additional
mass-squared contribution of the form $(m^2+n^2)R^{-2}$.  Therefore,
if the negative mass-squared term in the potential
 is smaller in magnitude than $R^{-2}$, then none of the higher Higgs KK
modes would acquire vevs.  We will assume this is the case in our
calculations, and the only fields that acquire vevs are
$\phi_u^{0\smbox{(00)}}$ and $\chi_{\tbox{R}}^{0,\smbox{(00)}}$, the zero-modes of
neutral Higgs fields.

Here we will only consider the  details of those gauge bosons in the
$(11)$ KK modes, and in this subsection it is understood  that we
have the superscript $(11)$. That is, we do not consider the $(01)$
and $(10)$ modes of $W^{\pm}_{R,\mu,5,6}$.
For a compact notation that will be convenient later on, for the
scalar partners ($G_5$ and $G_6$) of a generic vector gauge boson
($G_{\mu}$), we form the combinations
\begin{align}
G_{(\pm)}\equiv\frac{1}{\sqrt{2}}(G_5\pm G_6).
\end{align}
In the absence of Higgs mechanism, $G_{(+)}$ will be eaten by
$G_{\mu}$ at the corresponding KK-level, while $G_{(-)}$ will be
left as a physical degree of freedom.
Qualitatively,
$W^{\pm}_{R,\mu}$ eats a linear combination of $W^{\pm}_{R,(+)}$,
$W^{\pm}_{R,(-)}$, and $\chi_{\tbox{R}}^-$  (all fields with the superscript
$(01)$ and $(10)$), while the two remaining orthogonal directions
are left as physical degrees of freedom.

At the $(11)$-level, before symmetry breaking, we have the modes
\begin{align}
\mbox{Neutral Gauge Bosons:}&\quad W^{3}_{L,\mu},W^3_{R,\mu},B_{\mu}\nonumber\\
\mbox{Neutral Scalars:}&\quad W^3_{L,(+)}, W^3_{L,(-)},
W^3_{R,(+)},
W^3_{R,(-)},B_{(+)},B_{(-)},\phi_u^0,\chi_{\tbox{R}}^0\nonumber\\
\mbox{Charged Gauge Bosons:}&\quad W^{\pm}_{L\mu}\nonumber\\
\mbox{Charged Scalars:}&\quad
W^{\pm}_{L,(+)},W^{\pm}_{L,(-)},\phi_u^-.
\end{align}
Three (two) linear combinations of the neutral (charged) scalars
would be eaten, leaving seven (four) degrees of freedom (note that
the Higgs fields are complex). Since only the zero-mode Higgs
acquire vevs, the Higgs mechanism contribution to the mass matrix of
the neutral gauge bosons is same as that in Eq.~\ref{eq:HHWW}, and
we have an additional contribution of $2R^{-2}\mathbf{1}_{3\times
3}$.  We can diagonalize the (mass)$^2$ matrix up to
$\mathcal{O}(v_w^2/v_{\tbox{R}}^2)$ using the same unitary matrix $U_G$ and
obtain the eigenvalues to be those in Eq.~\ref{eq:gaugeeigen} with
the additional $2R^{-2}$.

Of the neutral scalars, we have several sets of particles that do
not mix with members of other sets at tree level:
\begin{align}
\mbox{Set 1:}&\quad \tfrac{1}{\sqrt{2}}\mbox{Re}[\phi_u^0],
\tfrac{1}{\sqrt{2}}\mbox{Re}[\chi_{\tbox{R}}^0]\nonumber\\
\mbox{Set 2:}&\quad W^3_{L,(-)}, W^3_{R,(-)}, B_{(-)}\nonumber\\
 \mbox{Set 3:}&\quad W^3_{L,(+)}, W^3_{R,(+)}, B_{(+)},
\tfrac{1}{\sqrt{2}}\mbox{Im}[\phi_u^0],
\tfrac{1}{\sqrt{2}}\mbox{Im}[\chi_{\tbox{R}}^0].
\end{align}
The squared-mass of particles in Set 1 are simply $2R^{-2}$ in
addition to the squared-masses of corresponding particles at
$(00)$-modes.  The mass matrix of particles in Set 2 are exactly
that of the neutral gauge bosons, with a lightest mode of $A_{(-)}$
with mass $m_{A_{(-)}}=m_{A_{\mu}}=\sqrt{2}R^{-1}$. Three linear
combinations of particles in Set 3 are eaten, and the two remaining
particles have masses that will depend on the Higgs potential.
As is the case with the zeroth-modes, as long as these Higgs are heavier
than the lightest gauge bosons, the values of their masses will not affect
our results about the dark matter of the model.

\subsection{Spectrum of Matter Fields}
Because there is no yukawa coupling between the Higgs doublet
$\chi_{\tbox{R}}$ and matter, at tree level all
 mass terms arise from the momentum in the extra dimensions and $v_w$.
The structure of the yukawa couplings, with the $Z_2\times
Z_2^{\prime}$ orbifolding ensures that the zero-mode matter fields
have the SM spectrum. As for the higher modes, the mass terms
arising from the extra dimension connect the left- and right-handed
components of a 6D chiral fermion $\Psi_{\pm}$, where $\pm$ denotes
6D chirality. The mass terms arising from electroweak
symmetry-breaking, however, connects left- and right-handed
components of two different 6D chiral fields. Taking the electron as
an example, the mass matrix of the electron KK modes in the basis
$\{ e_{1L}\ e_{1R}\ e_{2L}\ e_{2R}\}$ (with $e_{1L}$ and $e_{2R}$
having zero modes) is
\begin{align}
\mathcal{M}_{e^{(1)}}&=
\begin{pmatrix}\overline{e}_{1L}&\overline{e}_{1R}&
\overline{e}_{2L}&\overline{e}_{2R} \end{pmatrix}
\begin{pmatrix}
0 & R^{-1} & 0 & y_e\tfrac{v_2}{\sqrt{2}} \\
R^{-1} & 0 & y_e\tfrac{v_2}{\sqrt{2}} &0  \\
0 & y_e\tfrac{v_2}{\sqrt{2}} & 0 & -R^{-1}\\
y_e\tfrac{v_2}{\sqrt{2}} & 0 & -R^{-1} & 0
\end{pmatrix}
\begin{pmatrix} e_{1L}\\ e_{1R}\\e_{2L}\\e_{2R} \end{pmatrix}\nonumber\\
&=\begin{pmatrix}\overline{e}_{1}&\overline{e}_{2} \end{pmatrix}
\begin{pmatrix}  R^{-1} & y_e\tfrac{v_2}{\sqrt{2}} \\
y_e\tfrac{v_2}{\sqrt{2}} & -R^{-1} \end{pmatrix}
\begin{pmatrix} e_{1}\\ e_{2}\end{pmatrix}.
\label{eq:mattermatrix}
\end{align}
Generalizing this, we see that the $(mn)$ modes have masses
\begin{align}
m_{f^{(mn)}}=\left(\frac{N^2}{R^2}\pm m^2_{f^{(00)}}\right)^{1/2},
\end{align}
where $N^2=m^2+n^2$ and $m^2_{f^{(00)}}=y_f^2\frac{v_w^2}{2}$ is the
zero-mode mass of the fermion.

\subsubsection{Possible Dark Matter Candidates}
In order to see the dark matter candidates in our model, we look at the
spectrum of the KK modes (see Fig.~\ref{fig:level}). There are two classes of KK modes
of interest whose stability is guaranteed by KK parity: the ones with
$(-,-)$ and $(\pm, \mp)$ $Z_2\times Z^{\prime}_2$ quantum numbers. The former
have mass $\sqrt{2}R^{-1}$ and the latter $R^{-1}$.  We see from Fig.~\ref{fig:level}
that the first class of particles are the first KK mode of the
hypercharge gauge boson $B_Y$ and the second are the right handed
neutrinos $\nu_{\tbox{2L, 2R}}$. The presence of the RH neutrino dark matter
makes the model predictive and testable as we will see quantitatively in
what follows. The basic idea is that $\nu_{\tbox{R}}$ annihilation proceeds
primarily via the exchange of the $Z^{\prime}$ boson. So as the $Z^{\prime}$ boson mass
gets larger, the annihilation rate goes down very fast (like
$M^{-4}_{Z^{\prime}}$) and the $\nu_{\tbox{2}}$'s overclose the Universe.
Also since there
are lower limits on the $Z^{\prime}$ mass from collider searches \cite{zprime},
the $\nu_{\tbox{2}}$'s contribute a minimum amount to the $\Omega_{\tbox{DM}}$.
This leads to a two-component picture of dark matter and also adds
to direct scattering cross section making the dark matter detectable.
Below we make these comments more quantitative and present our detailed
results.

%
\section{Dark Matter Candidate I: $\nu_{\tbox{2L,2R}}$}
\label{sec:DM1}
\subsection{Annihilation Channels of $\nu_{\tbox{2L,2R}}$}
Since the yukawa couplings are small, except for the top-quark
coupling, we only consider annihilations through gauge-mediated
processes. For completeness, we first list the couplings between
matter fields and the neutral vector gauge bosons. For matter fields
charged under $SU(2)_1$, we have
\begin{align}
\mathcal{L}^{SU(2)_1}_{\overline{f}fB}&=
(\overline{q}\gamma^{\mu}P_Lq)\left[
\left(T^3_L+\frac{Y_{BL}}{2}\right)
\left(\frac{\gL\gR\gBL}{\gdenomA}\right)A_{\mu}
\right.\nonumber\\
&\quad\qquad\qquad
 +\left(\frac{T^3_L\gL^2(\gR^2+\gBL^2)-
\tfrac{Y_{BL}}{2}\gR^2\gBL^2}{\gdenomA\gdenomB}\right)Z_{\mu}
\nonumber\\
&\left.\quad\quad\qquad
+\left(\frac{Y_{BL}}{2}\frac{\gBL^2}{\gdenomB}\right)Z^{\prime}_{\mu}\right].
\end{align}
And for matter fields charged under $SU(2)_2$,
\begin{align}
\mathcal{L}^{SU(2)_2}_{\overline{f}fB}&=
(\overline{q}\gamma^{\mu}P_{\tbox{R}}q)\left[
\left(T^3_R+\frac{Y_{BL}}{2}\right)
\left(\frac{\gL\gR\gBL}{\gdenomA}\right)A_{\mu}
\right.\nonumber\\
&\quad\qquad\qquad +\left(-T^3_R-\frac{Y_{BL}}{2}\right)
\left(\frac{\gR^2\gBL^2}{\gdenomA\gdenomB}\right)Z_{\mu}
\nonumber\\
&\left.\quad\qquad\qquad
+\left(\frac{-T^3_R\gR^2+\tfrac{Y_{BL}}{2}\gBL^2}{\gdenomB}\right)
Z^{\prime}_{\mu}\right],
\end{align}
where $T^3_{\tbox{L}}=\pm\tfrac{1}{2}$ and $T^3_{\tbox{R}}=\pm\tfrac{1}{2}$
are the quantum number for the $SU(2)_1$ and $SU(2)_2$ groups respectively.
We choose this notation because $SU(2)_1$ is to be identified with $SU(2)_L$
of the Standard Model, even though there are right-handed particles
that are charged under the $SU(2)_1$ group.
Also, $Y_{\smbox{BL}}=+1/3$ for quarks and $Y_{\smbox{BL}}=-1$ for
leptons.

Using these formulas, the gauge interaction of the dark matter
candidates $\nu_{2L,2R}$ is given by the six-dimensional Lagrangian
\begin{align}
\mathcal{L}_{\nu}
&=-\frac{1}{2}(\overline{\nu}\gamma^{\mu}\nu)
\frac{\gR^2+\gBL^2}{\sqrt{\gBL^2+\gR^2}}Z^{\prime}_{\mu}.
\end{align}
We first notice that $\nu_{\tbox{2L,2R}}$ couple as a Lorentz vector.  Second,
we see that $\nu_{\tbox{2L,2R}}$ do not couple to $A_{\mu}$ nor $Z_{\mu}$
as expected because $\nu_{\tbox{2L,2R}}$ are singlets under the SM gauge
group. There is a small coupling between $\nu_{\tbox{2L,2R}}$ and $Z_{\mu}$
due to $Z^{\prime}_{\mu}-Z_{\mu}$ mixing. For the purpose of
evaluating annihilation cross sections, we can safely ignore this
mixing, as we will show.  However, this mixing will be important when we consider the direct
detection of $\nu_{\tbox{2L,2R}}$. In addition, we have the charged-current
interaction, similar to the SM case
\begin{align}
\mathcal{L}_{\smbox{CC}}=\frac{\gR}{\sqrt{2}}
\left(\overline{\nu}_2\gamma^{\mu}W_{2,\mu}^+P_{\tbox{R}}e_2+
\overline{e}_2\gamma^{\mu}W_{2,\mu}^-P_{\tbox{R}}\nu_2\right).
\end{align}
Even though $e_2$ is a Dirac spinor, its left-handed component has
$Z_2\times Z_2^{\prime}$ charge of $e_{2L}(--)$, and the
annihilation is kinematically forbidden.

%
%

\FIGURE{
\includegraphics[width=2in]{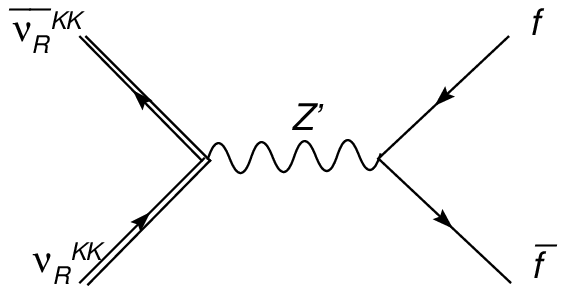}\quad\quad
\includegraphics[width=2in]{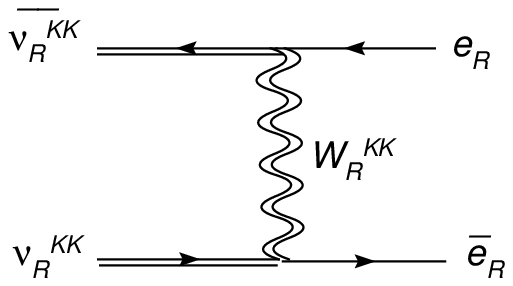}
\caption{Diagrams of annihilation channels of $\nu_{\tbox{2L,2R}}^{\tbox{KK}}$
to SM fermion-antifermion pairs.}
\label{fig:nu-anni}
}
Although we have two independent Dirac fermions for dark matter,
$\nu_{2}^{(10)}$ and $\nu_{2}^{(01)}$, they couple the same way to
$Z^{\prime}_{\mu}$ and have the same annihilation channels.  The
only difference is that, for charged current processes,
 $\nu^{(01)}$ ($\nu^{(10)}$) couples to $W_{2,\mu}^{\pm,(01)}$
($W_{2,\mu}^{\pm,(10))}$). The dominant contribution to the total
annihilation cross section of $\nu_{2L,R}$ is $s$-channel process
mediated by $Z^{\prime}_{\mu}$, as shown in Fig.\ \ref{fig:nu-anni}.
The thermal-averaged cross
section for $\langle\sigma(\overline{\nu}_2\nu_2\rightarrow
\overline{f}f)v_{\smbox{rel}}\rangle$, where $f$ is any chiral SM fermion
except the right-handed electron $e_{\tbox{R}}$, is
\begin{align}
\sigma(\overline{\nu}_2\nu_2\rightarrow
\overline{f}f)v_{\smbox{rel}} =
\frac{g^2_{(\overline{\nu}{\nu}Z^{\prime}_{\mu})}
g^2_{(\overline{f}{f}Z^{\prime}_{\mu})}}{12\pi}
\frac{s+2M_{\nu}^2}{(s-M^2_{Z^{\prime}})^2},
\end{align}
and with $s=4 M_{\nu}^2+M_{\nu}^2v^2_{\smbox{rel}}$, we expand in
$v^2_{\smbox{rel}}$,
\begin{align}
\sigma(\overline{\nu}_2\nu_2\rightarrow
\overline{f}f)v_{\smbox{rel}}
=\frac{g^2_{(\overline{\nu}{\nu}Z^{\prime}_{\mu})}
g^2_{(\overline{f}{f}Z^{\prime}_{\mu})}}{2\pi}
\frac{M_{\nu}^2}{(4M_{\nu}^2-M^2_{Z^{\prime}})^2}
\left[1+v_{\smbox{rel}}^2\left(\frac{1}{6}-\frac{2M_{\nu}^2}
{4M_{\nu}^2-M^2_{Z^{\prime}}}\right)\right].
\label{eq:S-general}
\end{align}

For the final state $\overline{e}_{\tbox{R}}{e}_{\tbox{R}}$, we have a $t$-channel process
through charged-current in addition to the $s$-channel
neutral-current process (see Fig. \ref{fig:nu-anni}). The cross-section therefore involves three
pieces: two due to the $s$ and $t$ channels and another from the
interference, denoted by $\sigma_{ss}$, $\sigma_{tt}$ and
$\sigma_{st}$ respectively.  Of these, $\sigma_{ss}$ has the same
form as Eq.~\ref{eq:S-general}, and we have
\begin{align}
\sigma(\overline{\nu}_2\nu_2\rightarrow
\overline{e}_{\tbox{R}}e_r)_{tt}v_{\smbox{rel}}
=\frac{\gR^4}{32\pi}\frac{M^2}{(M^2+M^2_{W_{\tbox{R}}})^2}
\left[1+v_{\smbox{rel}}^2\left(
\frac{3M^4+M^2M^2_{W_{\tbox{R}}}+M^4_{W_{\tbox{R}}}}{3(M^2+M^2_{W_{\tbox{R}}})^2}\right)\right],
\nonumber\\
\sigma_{st}v_{\smbox{rel}}
=\frac{g_{(\overline{\nu}{\nu}Z^{\prime}_{\mu})}g_{(\overline{e}_{\tbox{R}}{e}_{\tbox{R}}Z^{\prime}_{\mu})}\gR^2}
{4\pi(4M^2-M^2_{Z^{\prime}})(M^2+M^2_{W_{\tbox{R}}})}
\left[M^2-v^2_{\smbox{rel}}
\frac{M^2(40M^4+M^2_{W_{\tbox{R}}}M^2_{Z^{\prime}}+8M^2M^2_{W_{\tbox{R}}}-7M^2M^2_{Z^{\prime}})}
{12(4M^2-M^2_{Z^{\prime}})(M^2+M^2_{W_{\tbox{R}}})} \right]
\end{align}

\FIGURE{
\includegraphics[width=2in]{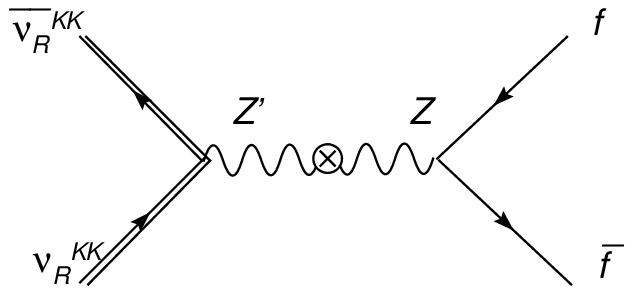}\quad\quad
\includegraphics[width=2in]{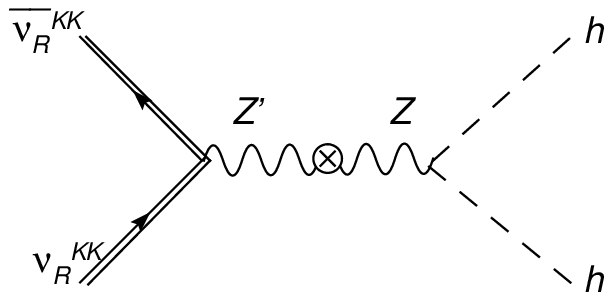}\\
\vspace{0.1in}
\includegraphics[width=2in]{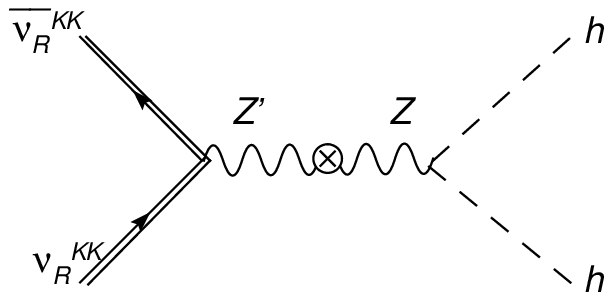}\quad\quad
\includegraphics[width=2in]{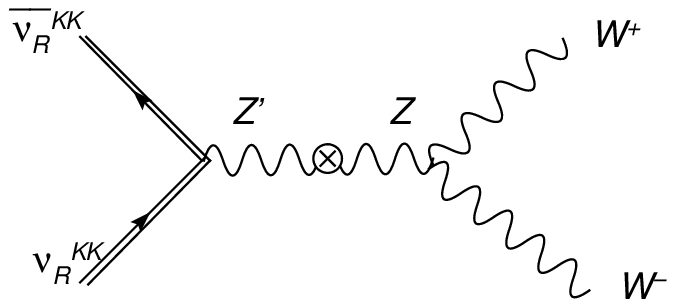}\quad\quad
\caption{Diagrams of annihilation channels of $\nu_{\tbox{2L,2R}}^{\tbox{KK}}$
to SM particles through $Z-Z^{\prime}$ mixing.}
\label{fig:nu-anni2}
}
Due to $Z-Z^{\prime}$, there can also be annihilation of KK neutrino
into SM Higgs, charged bosons, as well as fermion-antifermion pairs.
The diagrams for these processes are shown in Fig.\ \ref{fig:nu-anni2}.
In the limit that $v_w\ll v_{\tbox{R}}$,
we can work to the leading-order in the expansion of
$\mathcal{O}(v^2_w/v^2_{\tbox{R}})$, where we can estimate these processes by treating
the $Z-Z^{\prime}$ mixing as a
mass-insertion.
In terms of Feynman diagrams, these annihilation channels are $s$-channel processes,
where a pair KK neutrino annihilates into a $Z^{\prime}$-boson, which propagates
to the mixing vertex, converting $Z^{\prime}$ to $Z$, which then decays into
$h^{\ast}h$ (both neutral and charged), massless $W^+W^-$ or $\overline{f}{f}$.
Compared to the
amplitude of annihilation of KK neutrino into SM fermions without $Z-Z^{\prime}$ mixing, the
annihilation through mixing have effectively a replaced propagator
\begin{align}
\frac{1}{(s-M^2_{Z^{\prime}})}\rightarrow
\frac{1}{(s-M^2_{Z^{\prime}})} \delta\!M^2 \frac{1}{(s-M_Z^2)}
\end{align}
where
\begin{align}
\delta\!M^2\equiv\frac{\gR^2}{\sqrt{(\gL^2+\gY^2)(\gR^2+\gBL^2)}}M_Z^2,
\end{align}
is the off-diagonal element in the $Z-Z^{\prime}$ (mass)$^2$ matrix.
Since $s\sim 4M^2_{\nu}=4 R^{-2}$,  the annihilation cross section
into transverse gauge bosons and the Higgs bosons are suppressed by a factor of
$M_Z^4/s^2\sim (100 GeV)^4/ 16(500 GeV)^4\sim 10^{-4}$, and
can therefore be neglected.  The same is true for the annihilation to fermion-antifermion
pairs of the SM; we can ignore the effects of $Z-Z^{\prime}$ mixing in these channels.
As for the longitudinal modes, the ratio of annihilation cross-sections
of the longitudinal modes of the gauge bosons to the one single mode of SM fermion-antifermion pair is roughly
\begin{align}
\frac{\sigma(\nu^{\tbox{KK}}\nu^{\tbox{KK}}\rightarrow W^+W^-)}
{\sigma(\nu^{\tbox{KK}}\nu^{\tbox{KK}}\rightarrow \overline{f}{f})}
\sim  \left(\frac{\delta M^2}{m_W^2}\right)^2.
\end{align}
This ratio is about $\tfrac{1}{2}$ for $\gR=0.7 \gL$.  As there is only one annihilation mode into
the longitudinal modes of the charged gauge bosons, whereas there are many annihilation channels
to the SM fermion-antifermion pairs, the total annihilation cross section is dominated
by the SM fermion-antifermion contributions.

\subsection{Co-annihilation Contributions to the Relic Density of $\nu_{\tbox{2L,2R}}$}
In the MUED model, the KK mode of the left-handed electron, $e_{\smbox{L}}^{\tbox{KK}}$, is expected to be nearly
degenerate with the KK mode of the left-handed neutrino.  The self- and co-annihilation contribution
of $e_{\smbox{L}}^{\tbox{KK}}$ has been studied in the literature \cite{tait} \cite{kong}, where it is shown
that including such effects do not significantly alter the qualitative results, and that
$\nu_{\smbox{L}}^{\tbox{KK}}$ with a slightly different mass can still account for the observed relic density.
(However, $\nu_{\smbox{L}}^{\tbox{KK}}$ is ruled out by the direct detection experiments.  This
will be discussed in detail in Section \ref{sec:DirectD}.)

For our current model, the story is different.  As can be seen in Eq.~\ref{eq:leptons},
$e_{\tbox{2L,2R}}$,
the partners
of $\nu_{\tbox{2L,2R}}$ under $SU(2)_2$, carry different quantum numbers under
the $Z_2\times Z_2^{\prime}$ orbifold, and thus do not have $(10)$ nor $(01)$ modes.
There are states that are nearly degenerate with $\nu_{\tbox{2L,2R}}$, such as
the $e^{\prime}$ states.  However, these states interact with $\nu_{\tbox{2L,2R}}$ only
through Yukawa interactions, which can be ignored.  Therefore, we expect effects of self- and
co-annihilation with $\nu_{\tbox{2L,2R}}$ nearby states to be even smaller than
the MUED case, and ignore all such effects in our analysis.

\subsection{Important Differences in Comparison to Standard Analysis}
We note here that our our analysis
of the annihilation channels for $\nu^{\tbox{KK}}_{\tbox{2L,2R}}$
differ from those of \cite{tait} and
\cite{feng} for $\nu^{\tbox{KK}}_{\smbox{L}}$ in
two important ways.
First, in their analysis, the $s$-channel process is mediated by
$Z$-boson of the SM, whose mass can be ignored, whereas we have
$s$-channel processes mediated by $Z^{\prime}$, whose mass is
significantly larger than the mass of our dark matter candidate in
the region of interest.  Second, to a good approximation we can
discard $t$,$u$-channel processes mediated by charged gauge bosons
$W_2^{\pm}$, because $m^2_{W_2^{\pm}}$ has contributions both from
$R^{-1}$ and $v_{\tbox{R}}$.  To see this, let us make the approximation
$m^2_{W^{\pm}}=m^2_{Z^{\prime}}+R^{-2}$, then we compare the cross
section involving the product of a $t$ or $u$ diagram with a
$s$-channel diagram $\sigma_{st}$  with that coming from the square
of an $s$-channel diagram $\sigma_{ss}$,
\begin{align}
\frac{\sigma_{ss}}{\sigma_{st}}\approx
\frac{m^2_{\nu}+m^2_{W^{\pm}}}{4 m^2_{\nu}-m^2_{Z^{\prime}}}
=\frac{{2 (R^{-1})^2+m^2_{Z^{\prime}}}}{4
(R^{-1})^2-m^2_{Z^{\prime}}}.
\end{align}
Then $\sigma_{ss}\gg\sigma_{st}$ would require that
\begin{align}
\frac{{2 (R^{-1})^2+m^2_{Z^{\prime}}}}{4
(R^{-1})^2-m^2_{Z^{\prime}}}\gg 1 \quad\rightarrow\quad
m^2_{Z^{\prime}}\gg 2(R^{-1})^2,
\end{align}
 which is satisfied in the region of interest in the parameter space.
Similarly, the cross section involving two $t-$ or $u$-channel
diagrams, $\sigma_{tt},\sigma_{uu}$ or $\sigma_{tu}$ is small
compared to $\sigma_{ss}$.

\section{Dark Matter Candidate II: $\Bsca$ or $\Bvec$}
\label{sec:DM2}
The lightest (11) mode is either $\Bsca$ or $\Bvec$, depending on
radiative corrections.  Although in Reference \cite{ponton} found
that $\Bvec$ is heavier than $\Bsca$, this result is specific the
choice of orbifold in that particular case, and may not apply to
$Z_2\times Z_{2}^{\prime}$ orbifold that we have here. Instead of
performing the radiative corrections to determine which of the two
particles is lighter, we will do a phenomenological study exploring
both of these cases. To simplify the notation, we will often discard
the $(11)$ superscript in the fields.

\subsection{(Co)-Annihilation Channels of $\Bvec$}
When $v_w\ll R^{-1}$, $\Bvec$ is same as $A^{\tbox{(11)}}_{Y,\mu}$,
the KK mode of the photon up to small mixing effects.  The
annihilation channels and cross sections of $\Bvec$ have been
studied in detail in \cite{tait} and \cite{kong}, and in this
subsection we summarize their results.

\FIGURE{
\includegraphics[width=2in]{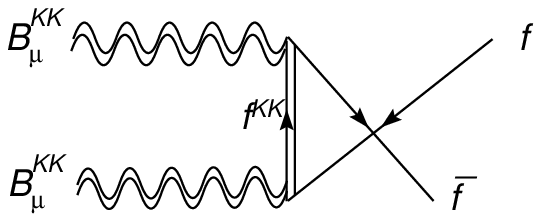}
\includegraphics[width=2in]{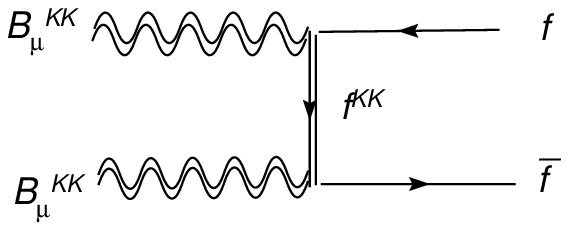}
\caption{Annihilation channels of a pair of $\Bvec$ into SM fermion-antifermion
pair.}
\label{fig:bvec-anni}
}

$\Bvec$ can annihilate itself into a fermion-antifermion pair
through $t$- and $u$-channel processes mediated by the (11) mode of
the fermion (Fig.\ \ref{fig:bvec-anni}).  It is important to note that the left- and
right-handed fermions of SM have separate massive KK modes with
vector-like couplings to the zero-mode fermion and $\Bvec$. The
annihilation cross section can be written as
\begin{align}
\sigma(\Bvec\Bvec\rightarrow\overline{f}f)
=g_1^4(Y_L^2+Y_{\tbox{R}}^4)N_c\frac{10(2
M_f^2+s)\mbox{ArcTanh}(\beta)-7s\beta}{72\pi s^2\beta^2},
\end{align}
where $M_f=\sqrt{2}R^{-1}$ is the mass of the KK-fermion exchanged,
$N_c$ is the color factor in the final state (3 for quarks and 1 for
leptons), and $Y_{L,R}$ is the hypercharge of the left- and
right-handed fermion.  Summing over all SM fermions gives
\begin{align}
\sum_{f\in
\tbox{SM}}N_c(Y_L^4+Y_{\tbox{R}}^4)=3(Y^4_{e_L}+Y^4_{e_{\tbox{R}}}+Y^4_{\nu_L}
+3(Y^4_{u_L}+Y^4_{u_{\tbox{R}}}+Y^4_{d_L}+Y^4_{d_{\tbox{R}}}))=\frac{95}{18}.
\end{align}

There are also annihilation channels to Higgs through $t$- and
$u$-channel processes mediated by a (11) mode of the Higgs boson as
well as a quartic interaction.  The annihilation cross section is
given by
\begin{align}
\sigma(\Bvec\Bvec\rightarrow
h^{\ast}h)=\frac{g_1^4Y^4_{\phi}}{6\pi\beta s},
\end{align}
where $Y_{\phi}=1/2$ is the hypercharge of the Higgs doublet. By
summing over two complex Higgs doublets, we have taken into account
the annihilation into the longitudinal zero modes of the $W$ and $Z$
gauge bosons.

In the MUED, the KK mode with mass closest to
$B^{\tbox{(1)}}_{\mu}$ is the KK mode of the right-handed
electron $e^{(1)}_{\tbox{R}}$ when radiative corrections are included
\cite{cheng1}. However, compared to the case without
co-annihilation, the qualitative results of the relic density due to
$B^{\tbox{(1)}}_{\mu}$ remains the same when one includes the
co-annihilation $e^{(1)}_{\tbox{R}} B^{\tbox{(1)}}_{\mu}\rightarrow
e_{R}A_{\mu}$ \cite{tait}. As pointed out by \cite{tait}, this is
because there are only two channel of such co-annihilation, leading
to a small co-annihilation cross section, and thus small change in
the relic density for a fixed $R^{-1}$.

In our case, we expect $\Bsca$ (which has no MUED analog) to be
close in mass to $\Bvec$ in addition to $e^{(11)}_{1R}$ (the analog
of $e^{(1)}_{\tbox{R}}$ in MUED).  Furthermore, the co-annihilation
$\Bvec\Bsca\rightarrow XX$ is significant as $\Bvec\Bsca$ can
annihilate to all SM fermions through $t$- and $u$-channel processes
mediated by a KK fermion.  The co-annihilation cross section to
fermion-antifermion pair is
\begin{align}
\sum_{f\in \tbox{SM}}\sigma(\Bvec\Bsca\rightarrow\overline{f}f)=
g_1^4\frac{95}{18}\frac{\mbox{ArcTan}(\beta)}{12\pi s \beta^2}.
\end{align}

Although the co-annihilation effect was overlooked in \cite{hsieh},
the most important conclusions of our previous work remain the same,
as we will show later.

\subsection{(Co)-Annihilation Channels of $\Bsca$}
The coupling of $\Bsca$ to matter fields in the full 6D-Lagrangian
is given by (in four-component notation)
\begin{align}
\mathcal{L}^{\tbox{6D}} &=\frac{g_1
Y}{\sqrt{2}}B_{\tbox{Y,-}}\left[\Psibar_{-}(i\gamma_5-\mathbf{1})\Psi_{-}
+\Psibar_{+}(i\gamma_5+\mathbf{1})\Psi_{+}\right]\nonumber\\
&=\frac{g_1
Y}{\sqrt{2}}B_{Y,-}\left[(-i-1)\Psibar_{\tbox{-L}}\Psi_{\tbox{-R}}
+(i-1)\Psibar_{\tbox{-R}}\Psi_{\tbox{-L}}
+(-i+1)\Psibar_{\tbox{+L}}\Psi_{\tbox{+R}}
+(i+1)\Psibar_{\tbox{+R}}\Psi_{\tbox{+L}}\right].
\end{align}
In terms of KK-modes, $\Bsca$ will couple to fermion fields in
(00-fermion)(11-fermion) pairs, and its annihilation channels to
fermions will proceed through $t-$ and $u-$processes mediated by a
KK-fermion.  The annihilation cross section is
\begin{align}
\sum_{f\in \smbox{SM}}\sigma(\Bsca\Bsca\rightarrow \overline{f}f)=
g_1^4\frac{95}{18}\frac{2(2M_f^2+s)\mbox{ArcTan}(\beta)-3s\beta}{2\pi
s\beta^2}.
\end{align}
In the non-relativistic limit, this cross-section is $p$-wave suppressed.
There is also annihilation to a pair of Higgs bosons through
the quartic coupling
\begin{align}
\mathcal{L}^{\tbox{4D}}&=g_1^2 Y_H^2\Bsca\Bsca H^{\dag\tbox{(00)}}H^{\tbox{(00)}},
\end{align}
and this gives a cross section of
\begin{align}
\langle\sigma v_{\tbox{rel}}\rangle=\frac{g_1^4 Y_H^4}{2\pi s}.
\end{align}

Because the annihilation of $\Bsca$ to fermion modes is $p$-wave suppressed,
the relic density resulting $\Bsca$ self-annihilation channels would in general
be too high. Therefore, we must rely on co-annihilation channels such as $\Bvec\Bsca\rightarrow XX$
to obtain observed relic density, as we will see in the next section.

\section{Numerical Results of Relic Density}
\label{sec:RelicResult}
The main free parameters of our theory are $R^{-1}$ and
$M_{Z^{\prime}}$, and the mass-splitting
$\Delta\equiv(M_{\Bvec}-M_{\Bsca})/M_{\Bsca}$.  In addition, we have
$\gR$ or $\gBL$ as a free parameter as long as we can satisfy the
constraint
\begin{align*}
g_1^2=\frac{\gBL^2\gR^2}{\gBL^2+\gR^2}.
\end{align*}

\subsection{$\Bvec$-$\nu^{\tbox{(01)}}_{\tbox{2L,2R}}$ Dark Matter}
\FIGURE{
\includegraphics[width=2.75in]{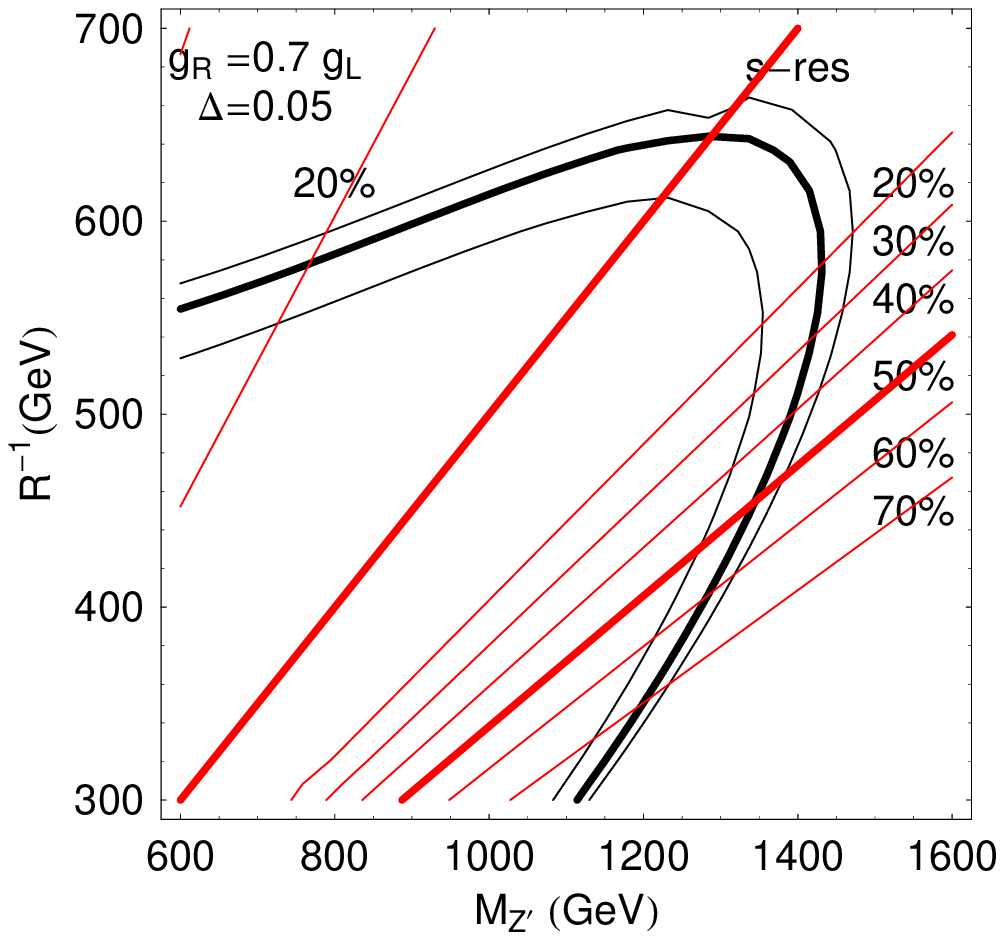}
\includegraphics[width=2.75in]{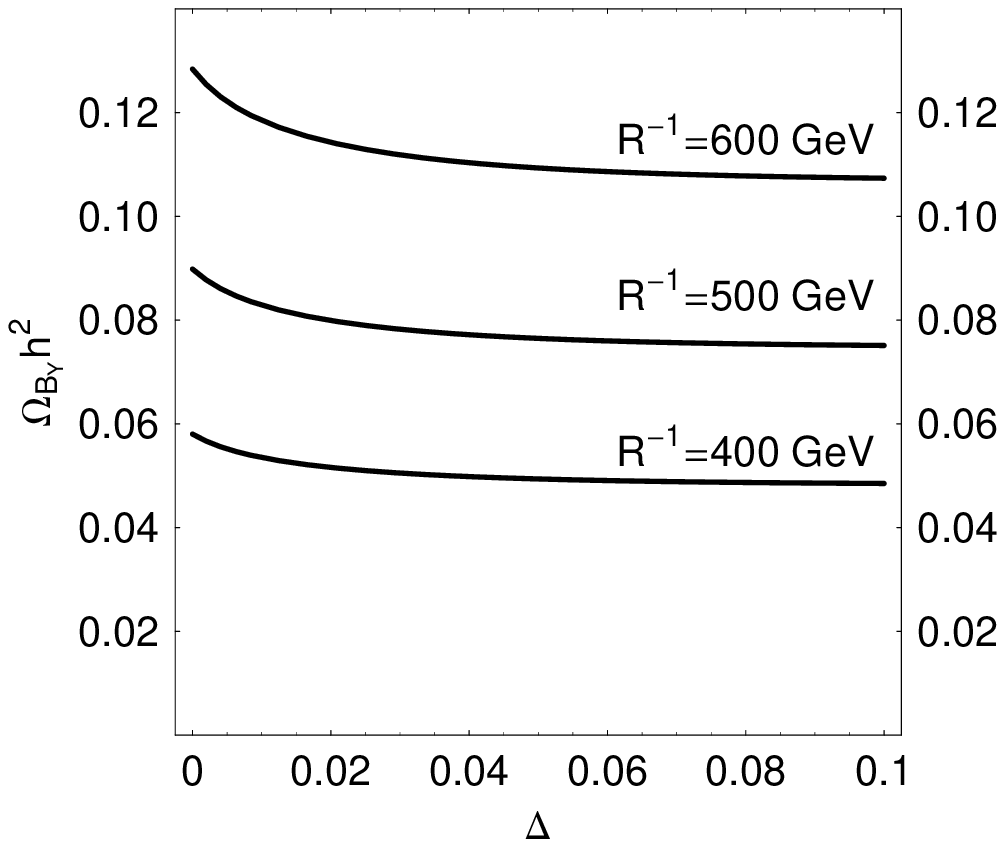}
\caption{The plot on the left shows the contour in the $R^{-1}-M_{Z'}$ plane that corresponds to
$\Omega_{\nu_{L,R}}h^2 + \Omega_{B_Y}h^2$  being the observed dark
matter. The intersection of the red lines with the contour indicate
the the fraction of KK neutrinos in the dark matter.
The plot on the right shows $\Omega_{B_Y}h^2$ as a function of $\Delta$ for
various values of $R^{-1}$.}
\label{fig:MainPlotVector}
}

For $\Delta=0.05$, we present the allowed region in
$R^{-1}-M_{Z^{\prime}}$ space that gives the observed dark
matter relic density in the first plot of Fig.~\ref{fig:MainPlotVector}.
Since both
$\Bvec$ and $\nu^{\tbox{(01)}}_{\tbox{2L,2R}}$ can independently give the
correct relic density without co-annihilation from other modes with
almost degenerate mass, varying $\Delta$ does not affect the
qualitative results of what we present below.
The independence of $\Omega_{\Bvec}h^2$ on $\Delta$ as can
be seen in the second plot of Fig.~\ref{fig:MainPlotVector}.
For small values of
$M_{Z^{\prime}}$, the annihilation of $\nu^{\tbox{(01)}}_{\tbox{2L,2R}}$
is efficient and most of the dark matter is $\Bvec$ having a mass of
roughly $\sqrt{2}R^{-1}\sim 700$ GeV.  In fact, along the line
$2M_{\nu^{(01)}}=2R^{-1}=M_{Z^{\prime}}$, the annihilation of
$\nu^{\tbox{(01)}}_{\tbox{2L,2R}}$ has an $s$-channel resonance, and its
contribution to dark matter relic density is minimal.  Away from the
line of $s$-channel resonance, the contribution of
$\nu^{\tbox{(01)}}_{\tbox{2L,2R}}$ to the relic density increases, and
$R^{-1}$ decreases so as to decrease the relic density due to
$\Bvec$, keeping the total relic density within the allowed range.

The current experimental bound on the massive, neutral, vector boson
is $M_{Z^{\prime}}>800$ GeV.  If we further impose the bound that
$R^{-1}>400$ GeV, the allowed region in the parameter space is very
limited.

\subsection{$\Bsca$-$\nu^{\tbox{(01)}}_{\tbox{2L,2R}}$ Dark Matter}
\FIGURE{
\includegraphics[width=3in]{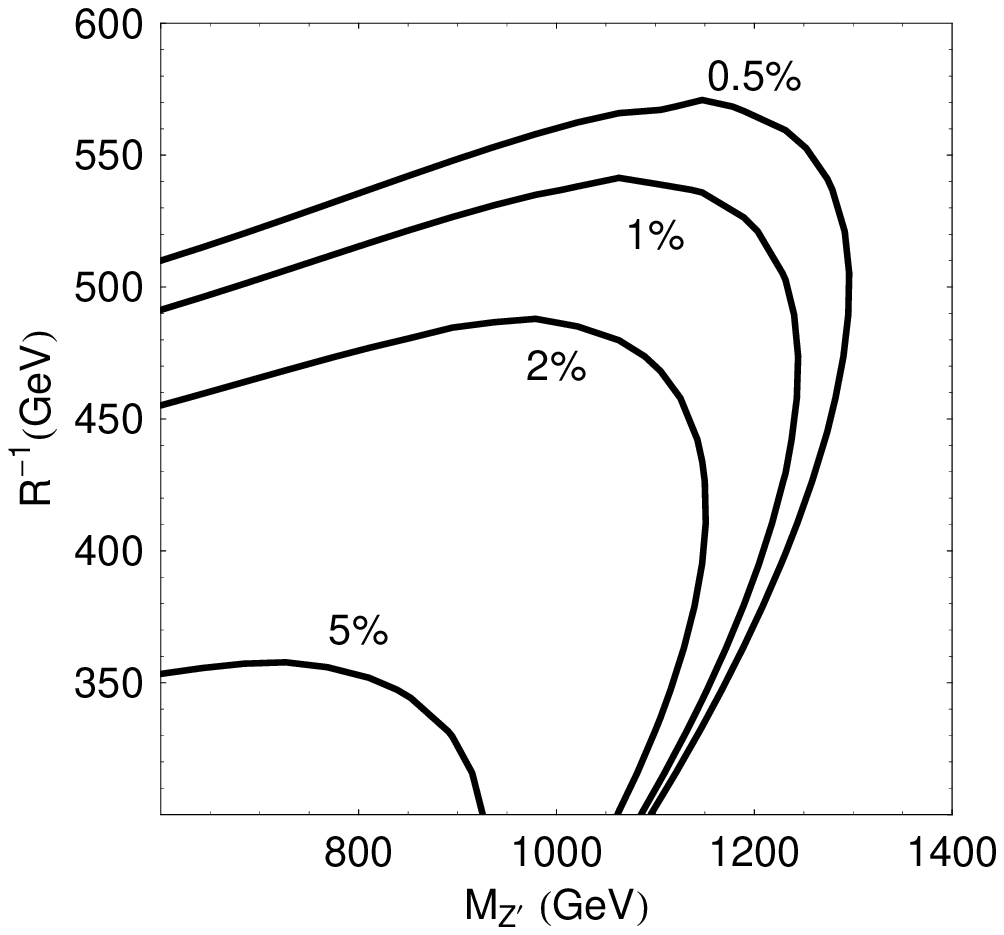}
\includegraphics[width=2.75in]{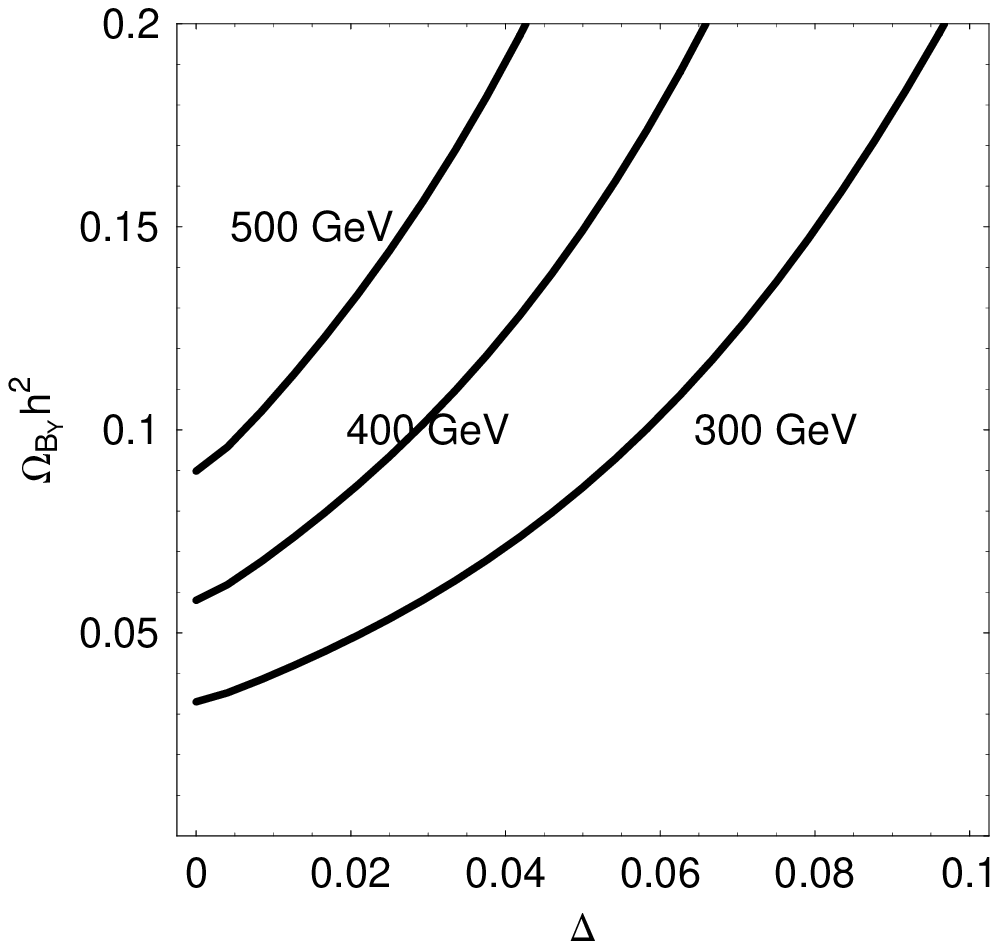}
\caption{The plot on the left shows the allowed region in the
parameter space that gives rise to the observed dark matter relic
density for $\gR=0.7\gL$ and different values of $\Delta$. On the
right, we plot the relic density due to $\Bsca$ as a function of the
mass-splitting $\Delta$ for various values of $R^{-1}$.}
\label{fig:MainPlotBM}
}

As stated earlier, $\Bsca$ by itself can not annihilate efficiently
enough to account for the observed relic abundance.  However, there
is significant co-annihilation process $\Bsca\Bvec\rightarrow
\overline{f}f$.  In Fig.~\ref{fig:MainPlotBM}, we show contours that
give the observed relic density for various values of $\Delta$. We
see that when $\Bsca$ and $\Bvec$ are nearly degenerate to less than
5\%, then the distribution of dark matter among
$\nu^{\tbox{(01)}}_{\tbox{2L,2R}}$ and $\Bsca$ is similar to the previous
case.  When the mass splitting between  $\Bsca$ and $\Bvec$ is
larger than 5\%, however, the model is ruled out as we can not
obtain the observed relic density without violating $R>400$ GeV
bound.  When $\Bsca$ and $\Bvec$ are nearly degenerate,
$\nu^{\tbox{(KK)}}_{\tbox{2L,2R}}$ can still contribute significantly to
the observed relic density when $M_{Z^{\prime}}$ is about $1.2$ TeV
and $R^{-1}\sim 400$ GeV.

\section{Direct Detection of Two-Component Dark Matter}
\label{sec:DirectD}
As we have a two-component dark matter, the total dark
matter-nucleon cross section is given by
\begin{align}
\sigma_n=\kappa_{\nu_{\tbox{R}}}\sigma_{\nu_{\tbox{R}}}+\kappa_B \sigma_B,
\end{align}
where $\sigma_{\nu_{\tbox{R}}(B)}$ is the spin-independent KK neutrino
(hypercharge vector or pseudoscalar)- nucleon scattering  cross section, and
\begin{align}
\kappa_{\nu_{\tbox{R}}}\equiv \frac{\Omega_{\nu_{\tbox{R}}} h^2 }{\Omega_{\nu_{\tbox{R}}} h^2
+\Omega_{B} h^2 },
\end{align}
is the fractional contribution of the KK neutrino relic density
to the total relic density of the dark matter.  $\kappa_B$ is
similarly defined.  As pointed out in Ref. \cite{tait},
$\sigma_B$ is of the order $\sigma_B\sim10^{-10}$ pb, and we will find that
$\sigma_{\nu_{\tbox{R}}}\gg\sigma_B$.  Therefore, it is a good approximation to
take $\sigma_n$ as
\begin{align}
\sigma_n\approx\kappa_{\nu_{\tbox{R}}}\sigma_{\nu_{\tbox{R}}}.
\end{align}

The elastic cross section between $\nu_{\tbox{2L,2R}}$ and a
nucleon inside a nucleus $N(A,Z)$ is given by
\begin{align}
\sigma_0&=\frac{b_N^2 m^2_n}{\pi A^2},
\end{align}
 where $b_N=Z b_p+(A-Z)b_n$ and $b_{p,n}$ is the effective four-fermion
coupling between $\nu_{\tbox{2L,2R}}$ and nucleon.   They are given by
$b_p=2b_u+b_d$ and $b_n=b_u+2b_d$.  In our case, although
$\nu_{\tbox{2L,2R}}$ only couples to $Z^{\prime}_{\mu}$ at leading order,
we have to taken into account the $Z-Z^{\prime}$ mixing. We can
including the effects of mixing up to order of
$\mathcal{O}(M^2_Z/M^2_{Z^{\prime}})$ by treating the mixing as
perturbations and include one vertex mixing. In this case, we have
\begin{align}
b_q&=\frac{1}{2M^2_{Z^{\prime}}}
g_{(\overline{\nu_2}{\nu_2}Z^{\prime})}
\left[(g_{(\overline{q}_L{q}_LZ^{\prime})}
+g_{(\overline{q}_{\tbox{R}}{q}_{\tbox{R}}Z^{\prime})}) -(g_{(\overline{q}_L{q}_LZ)}
+g_{(\overline{q}_{\tbox{R}}{q}_{\tbox{R}}Z)})\frac{\delta\!M^2}{M^2_Z}
+\mathcal{O}\left(\frac{M^2_Z}{M^2_{Z^{\prime}}}\right) \right],
\end{align}
where
\begin{align}
\delta\!M^2\equiv\frac{\gR^2}{\sqrt{(\gL^2+\gY^2)(\gR^2+\gBL^2)}}M_Z^2
\end{align}
is the mixing between $Z$ and $Z^{\prime}$ (see Eq.~\ref{eq:gaugematrix}).
%

The prospects of direct detection of $\Bvec$ has been studied
extensively, and the calculated detection rates are beyond the reach
of current experiments.  As for $\Bsca$, because there is no
$s$-wave for elastic scattering $\Bsca N \rightarrow\Bsca N$, the
cross-section is suppressed by a factor of $v_{\tbox{rel}}^2\sim 10^{-5}$.
Therefore, we expect that the direct detection rates of
$\nu^{(10)}_{\tbox{2L,2R}}$ will dominate that of both $\Bvec$ and $\Bsca$.
This is one of the main points of our work: the lightest KK-mode of
sterile neutrino as dark matter candidate could be detected directly
in the current and the next rounds of direct-detection experiments if its
relic density is significant compared to the observed total relic density,
in contrast to
other dark matter candidates in the literature, such as the
neutralino of MSSM or the lightest KK-mode of the photon.

\FIGURE{
\includegraphics[width=3in]{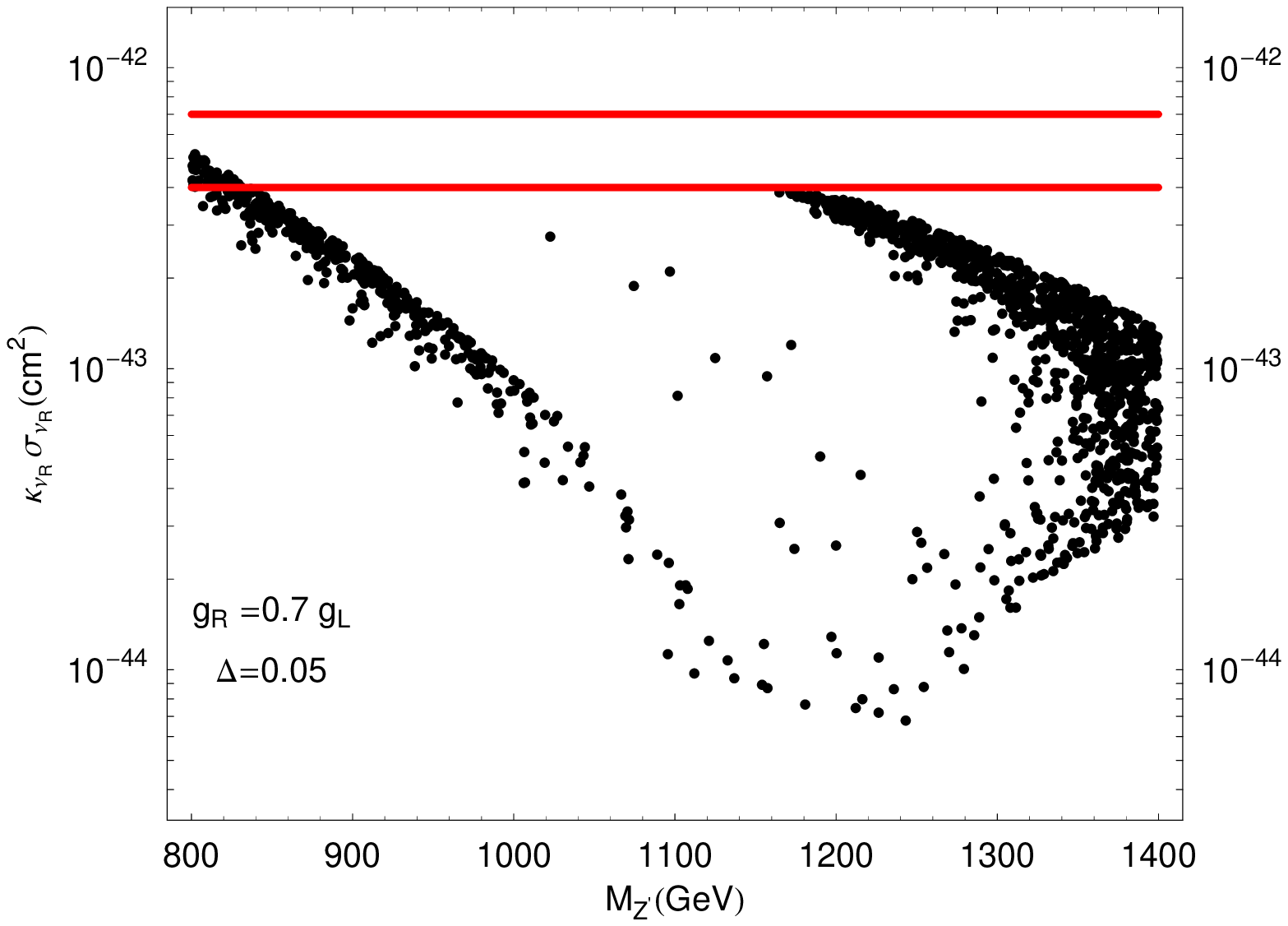}
\includegraphics[width=2.5in]{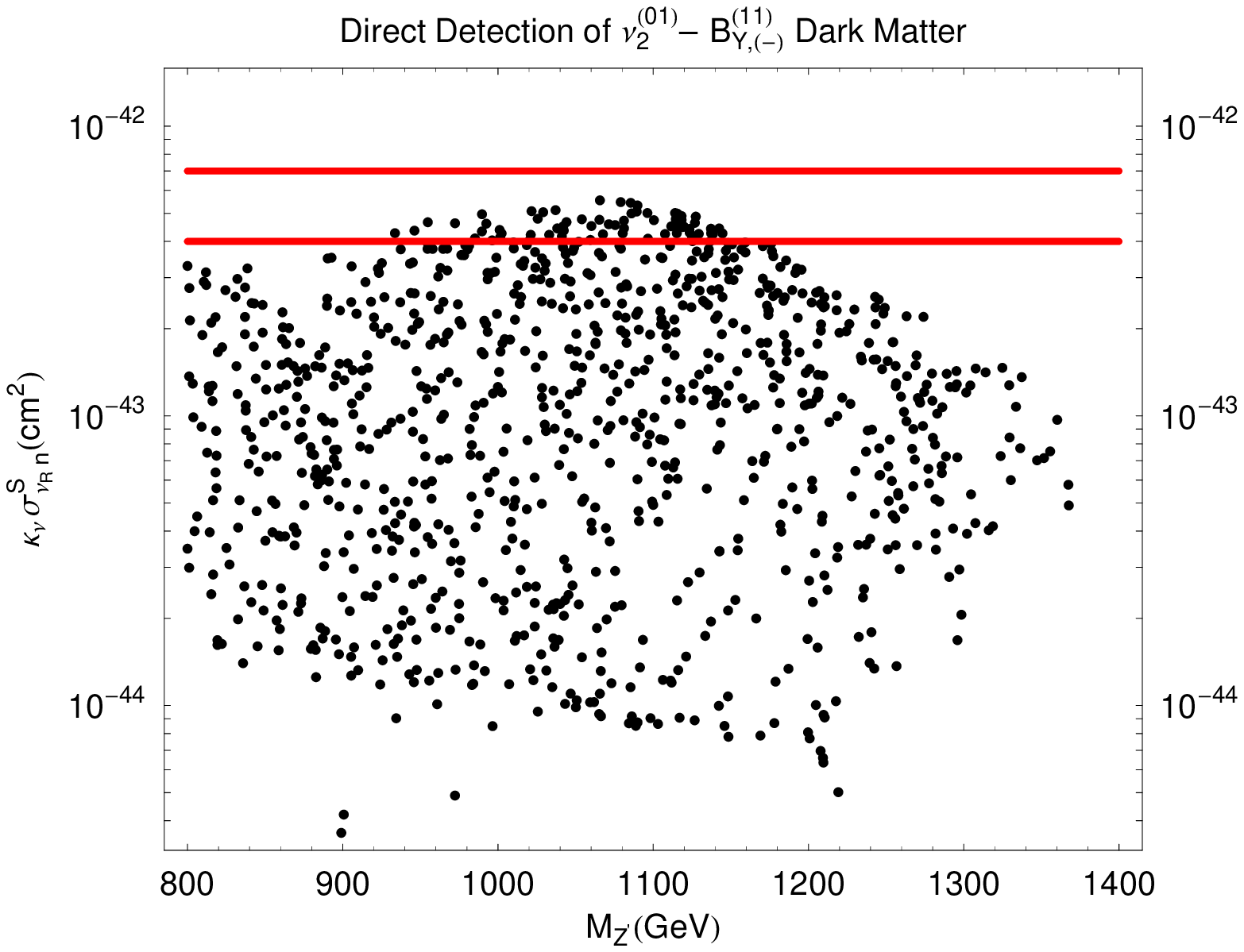}
\caption{The plot
on the left (right) shows the dark matter-nucleon cross-section as a
function of $M_{Z^{\prime}}$ for the case
where $\Bvec$ ($\Bsca$) is lightest (11) mode.  The plots
scan over different values of $R^{-1}$ and
$-0.05< \Delta <0$ that gives the observed relic density.
The horizontal lines correspond to the upper bounds on
$\sigma_n$ from CDMS II
for dark matter candidates with masses 300
and 500 GeV}
\label{fig:dd}
}

In Fig.~\ref{fig:dd}, we show the direct detection cross section as a function
of $M_{Z^{\prime}}$ for both cases where $\Bvec$ and $\Bsca$ is the lighter of the two.
The horizontal lines correspond to the upper bounds on
$\sigma_n$ from CDMS II
for dark matter candidates with masses 300
and 500 GeV, which are about 4$\times
10^{-43} \mbox{cm}^2$ and 7$\times
10^{-43} \mbox{cm}^2$, respectively.

A particularly interesting region in the parameter space
is $R\sim 400$ GeV and $M_{Z^{\prime}}\sim 1200$ GeV.  Here, the
KK-sterile neutrino contributes to roughly half of the relic
density.  This admixture of dark matter is just below the current
experimental bound from direct detection, as shown in Fig.
\ref{fig:MainPlotVector}, when we use the CDMS II bound that dark
matter-nucleon spin-independent cross-section must not exceed
$4\times 10^{-43} \mbox{cm}^2$ for a 400 GeV dark matter.

\section{Some Phenomenological Implications}
\label{sec:WZPheno}

In this section, we  give a qualitative comparision of the phenomenological
implications of this model with
those of the conventional left-right symmetric models \cite{lr}. It has
long been recognized that two important characteristic predictions of the
left-right models are the presence of TeV scale $W_R$ and $Z'$ gauge
bosons which can be detectable in high energy colliders \cite{kayser}.
In addition to the collider signatures of generic UED models \cite{macesanu},
two predictions characteristic of the model discussed here which differ
from those of the earlier models are : (i) The mass of $Z'$ has an upper
bound of about $1.5$ TeV and a more spectacular one where (ii) the $W_R$
in this model, being a KK
excitation, does not couple to a pair of the known standard model
fermions which are zero modes. This property of the $W_R$ has a major
phenomenological impact and will require a completely new analysis of
constraints on this e.g. the well known $K_L-K_S$ mass difference
constraint on $M_{W_R}$ \cite{soni} does not apply here since the mixed
$W_L-W_R$ exchange box
graph responsible for the new contribution to $K_L-K_S$ mass difference
does not exist. The box graph where both exchange particles are $W_R$'s
exists but its contribution to the $\Delta S=2$ Hamiltonian is suppressed
compared to the left-handed one by a factor
$\left(\frac{M_{W_L}}{M_{W_R}}\right)^4$ and gives only a very weak bound
on $M_{W_R}$.

Also, the bounds from muon and beta decay \cite{beg} are
nonexistent for the same reason because there is no tree
level $W_R$ contribution to these processes. Furthermore, in this model,
there is no $W_L-W_R$ mixing unlike the conventional left-right models.

Because of this property, the decay modes and production mechanism of the
$W_R$ are also very different from the case of the conventional
left-right model, while the decay modes and production mechanism of the
$Z'$ remain the same. We do not discuss the $Z'$ case which has been very
widely discussed in literature.

The $W_R$ will have a mass given by As far as the $W_R$ is concerned, it
is given by the formula $M^2_{W_R}\sim \frac{cos^2\theta_W}{cos
2\theta_W}M^2_{Z'}$. Furthermore, it can only be pair produced and will decay
to $u'_{\tbox{2L}}\bar{d'}_{\tbox{2L}},\ u_{\tbox{2R}}\bar{d}_{\tbox{2R}},\
u'_{\tbox{2R}}\bar{d'}_{\tbox{2R}},\ u_{\tbox{2L}}\bar{d}_{\tbox{2L}},
\ \bar{e}_{\tbox{2L,R}}\nu_{\tbox{2L,R}},$ and
$\bar{e'}_{\tbox{2L,R}}\nu'_{\tbox{2L,R}}$.  For sub-TeV $W_R$, only the decay modes
$\bar{e'}_{\tbox{2L}}\nu'_{\tbox{2L}}$ and $u_{\tbox{2R}}\bar{d}_{\tbox{2R}}$ will dominate
depending on the precise value of $W_R$ mass and the $R^{-1}$.
The leptonic decay mode will
look very similar to the supersymmetric case where pair-produced sleptons
will decay to a lepton and the neutralino. The hadronic channel will
however look different from the squark case.  The further details of the
collider signature of our model is currently under investigation and will
be presented separately.

\section{Conclusions}
\label{sec:Conclusions}
 In summary, we have studied the profile of cold dark matter candidates
in a Universal Extra Dimension model with a low-scale extra $W_{\tbox{R}}$
and $Z'$. There are two possible candidates: $\nu^{\tbox{KK}}_{\tbox{R}}$ and either
the $\Bsca$ or $\Bvec$ depending on which one receives less
radiative corrections. We have done detailed calculation of the
relic density of these particles as a function of the parameters of
the model which are $g_{\tbox{R}}$, $R^{-1}$ and $M_{Z'}$. We find upper
limits on these parameters where the above KK modes can be cold dark
matter of the universe. In discussing the relic abundance, we have
considered the co-annihilation effect of nearby states. We also
calculate the direct detection cross-section in current underground
detectors for the entire allowed parameter range in the model and we
find that, for the case where KK neutrino contributes
significantly to the total relic density,
the lowest possible value of the cross-section predicted
by our model is accessible to the current and/or planned direct
search experiments.  Therefore, the most interesting region of our model can
not only be tested in the colliders but also these dark matter
experiments. Combined with LHC search for the $Z'$ of left-right model,
dark matter experiments could rule out this model.

\section{Acknowledgement}
 The work of K.H and R.N.M is supported by the National Science Foundation grant
no. Phy-0354401. S.N is supported by the DOE grant no. Phy.
DE-FG02-97ER41029.

\appendix
\addcontentsline{toc}{section}{Appendices}
\section{Fields on $T^2/{Z_2\times Z'_2}$}
For convenience of type-setting, we define the functions
\begin{align}
c(i,j)\equiv \cos\frac{i x^5+j x^6}{R},\quad\quad
s(i,j)\equiv \sin\frac{i x^5+j x^6}{R},\nonumber\\
c^{\prime}(i,j)\equiv \cos\frac{i x^{\prime 5}+j x^{\prime 6}}{R},\quad\quad
s^{\prime}(i,j)\equiv \sin\frac{i x^{\prime 5}+j x^{\prime 6}}{R},
\end{align}
And for reference we will make use of this integral
\begin{align}
\int_0^{2\pi R}\!\!\!\!\!\!\!\! dx^5 \int_0^{2\pi R}\!\!\!\!\!\!\!\!
dx^6 c(i,j) c(m,n)
=\int_0^{2\pi R}\!\!\!\!\!\!\!\! dx^5
 \int_0^{2\pi R}\!\!\!\!\!\!\!\! dx^6
s(i,j) s(m,n)
=2\pi^2
R^2\delta_{im}\delta{jn}
\end{align}
for positive integers $i,j,m$ and $n$, extensively.
We have the compactified space $2\pi R\times 2\pi R$ by imposing the
periodic boundary conditions on the fields
\begin{align}
\phi(x^{\mu},x^4,x^5)=\phi(x^{\mu},x^4+2\pi
R,x^5)=\phi(x^{\mu},x^4,x^5+2\pi R).
\end{align}
The periodic boundary conditions mean that we can write the fields
in the form of
\begin{align}
\phi(x^{\mu},x^4,x^5)=\sum_{n,m}\left( c(n,m)\varphi^{(nm)}(x^{\mu})
+s(n,m)\tilde{\varphi}^{(nm)}(x^{\mu})\right).
\label{eq:KKexpand}
\end{align}
On top of the periodic boundary conditions, we impose two
orbifolding symmetries on our theory
\begin{align}
Z_2:{\bf{y}}\rightarrow -{\bf{y}},\quad
Z_2^{\prime}:{\bf{y^{\prime}}}\rightarrow -{\bf{y^{\prime}}}.
\end{align}
with ${\bf y^{\prime}}={\bf y}-(\pi R/2,\pi R/2)$.
Demanding that the Lagrangian be invariant under
the orbifolding symmetries, we can assign parities to the fields under
the discrete transformations and remove roughly half of the KK
modes in Eq.~\ref{eq:KKexpand}.  The choices of signs are motivated
by the desired phenomenology.  In our case, we have two orbifolding
symmetries, so we can assign two signs to a given field. There are
four possibilities: $(+,\pm)$ and $(-,\pm)$, and we examine each
case separately.

For $(+,\pm)$ case, we have the general expansion
\begin{align}
\phi^{(+,\pm)}(x^{\mu},x^4,x^5)&=\frac{1}{2\pi
R}\sum_{n,m=0}^{\infty}\left( c(n,m)\varphi^{(nm)}
\right)\nonumber\\
&=\frac{1}{2\pi R}\sum_{n,m=0}^{\infty}\left[ \left(c^{\prime}(n,m)
\cos\frac{(m+n)\pi}{2}- s^{\prime}(n,m)\sin\frac{(m+n)\pi}{2}\right)\varphi^{(nm)}\right].
\label{eq:KKexpand++}
\end{align}
So we see that for $(+,+)$ fields, we need $n+m$ and $n-m$ to be
even.

For $(+,-)$ fields, we need $n+m$ and $n-m$ to be odd. For
$(-,\pm)$ case, we have the general expansion
\begin{align}
\phi^{(-,\pm)}(x^{\mu},x^4,x^5)&=\frac{1}{2\pi
R}\sum_{n+m\geq1}^{\infty}\left( s(n,m)\varphi^{(nm)}
\right)\nonumber\\
&=\frac{1}{2\pi R}\sum_{n+m\geq1}^{\infty}\left[ \left(s^{\prime}(n,m)\cos\frac{(m+n)\pi}{2}+
c^{\prime}(n,m)\sin\frac{(m+n)\pi}{2}\right)\varphi^{(nm)}\right].
\label{eq:KKexpand--}
\end{align}
So we see that for $(-,+)$ fields, we need $n+m$ and $n-m$ to be
odd.  For $(-,-)$ fields, we need $n+m$ and $n-m$ to be even.  Of
course, for the $(-,\pm)$ cases, we can not have $(m,n)=(0,0)$ mode.

\section{Normalization of Fields and Couplings}
\subsection{Matter Fields}
The dark matter candidates of the theory are the first KK modes of
the neutrinos charged under $SU(2)_2$.  They have the $Z_2\times
Z_2^{\prime}$ charges: $\nu_{\tbox{2L}}(-,+),\nu_{\tbox{2R}}(+,-)$.  If we let
$\vect{n}=(n,m)$, we see each of $\nu_{\tbox{2L,2R}}$ has two independent modes:
$\vect{n}=(1,0)$ and $\vect{n}=(0,1)$.  These are two independent Dirac particles
in the sense that there is no mixing at tree level in the effective 4D theory.

We expand the kinetic energy term
\begin{align}
\mathcal{L}_{\tbox{6D-KE}}&=i\overline{\Psi}\Gamma^M\partial_M\Psi\nonumber\\
&=i\begin{pmatrix} \overline{\Psi}_- & \overline{\Psi}_+ \end{pmatrix}
\begin{pmatrix}0 & \gamma^{\mu}\partial_{\mu}+i\gamma_5\partial_5+\partial_6 \\
\gamma^{\mu}\partial_{\mu}+i\gamma_5\partial_5-\partial_6 &
0\end{pmatrix}
\begin{pmatrix}\Psi_+ \\ \Psi_- \end{pmatrix}\nonumber\\
&=i\overline{\Psi}_-(\gamma^{\mu}\partial_{\mu}+i\gamma_5\partial_5+\partial_6)\Psi_-
+i\overline{\Psi}_+(\gamma^{\mu}\partial_{\mu}+i\gamma_5\partial_5-\partial_6)\Psi_+.
\end{align}
Note that $\Psi$ is an eight-component object, while $\Psi_{\pm}$
are four-component, six-dimensional chiral spinors.  We denote
six-dimensional chirality by $\pm$ and four-dimensional chirality by
$L,R$.  Each six-dimensional chiral spinor is vector-like in the
four-dimensional sense, and each is a Dirac spinor.  Since our dark
matter candidate is of (-1) 6D-chirality, we only deal with the first part of
the kinetic energy term, and drop the subscript.

Since we are after the coefficients, we expand in detail the first
KK mode of the dark matter candidate.
\begin{align}
\mathcal{L}_{\tbox{6D-KE}}& \supset
i\overline{\Psi}_-(\gamma^{\mu}\partial_{\mu}+i\gamma_5\partial_5+\partial_6)\Psi_-\nonumber\\
&=i\begin{pmatrix} \Psi^{\dag}_L & \Psi^{\dag}_{\tbox{R}} \end{pmatrix}
\begin{pmatrix}i\partial_5+\partial_6 & \sigma^{\mu}\partial_{\mu} \\
\overline{\sigma}^{\mu}\partial_{\mu} &
-i\partial_5+\partial_6\end{pmatrix}
\begin{pmatrix}\Psi_{\tbox{R}} \\ \Psi_L \end{pmatrix}\nonumber\\
&=i(\Psi^{\dag}_L\sigma^{\mu}\partial_{\mu}\Psi_L+
\Psi^{\dag}_{\tbox{R}}\overline{\sigma}^{\mu}\partial_{\mu}\Psi_{\tbox{R}}
+\Psi^{\dag}_L(i\partial_5+\partial_6)\Psi_{\tbox{R}}+\Psi_{\tbox{R}}(-i\partial_5+\partial_6)\Psi_L)
\label{eq:nuKE}
\end{align}
At this point, we use the KK-expansions.  Noting the charge
assignments $\nu_{2L}(-,+),\nu_{2R}(+,-)$, we expand the fields as
\begin{align}
\nu_{\tbox{2R}}=\frac{1}{\sqrt{2}\pi R}\left(c(1,0)\nu^{\tbox{(10)}}_{\tbox{2R}}(x^{\mu})
+c(0,1)\nu^{\tbox{(01)}}_{\tbox{2R}}(x^{\mu})\right),\nonumber\\
\nu_{\tbox{2L}}=\frac{1}{\sqrt{2}\pi R}\left(s(1,0)\nu^{{\tbox{(10)}}}_{\tbox{2L}}(x^{\mu})
+i s(0,1)\nu^{\tbox{(01)}}_{\tbox{2L}}(x^{\mu})\right).
\label{eq:nuKKexpand}
\end{align}
The four-dimensional effective
Lagrangian is obtained by inserting the expansion of Eq.~\ref{eq:nuKKexpand}
into Eq.~\ref{eq:nuKE}, and integrate $x^5$ and
$x^6$ from $0$ to $2\pi R$.  Following this procedure, we obtain
\begin{align}
\mathcal{L}_{\tbox{4D-eff}}=
\int_0^{2\pi R}\!\!\!\!\!\!\!\! dx^5
\int_0^{2\pi R}\!\!\!\!\!\!\!\!  dx^6
\mathcal{L}_{\tbox{6D-KE}}&=
i\left(
 \nu_{\tbox{2L}}^{\dag\tbox{(10)}}\sigma^{\mu}\partial_{\mu}\nu_{\tbox{2L}}^{\tbox{(10)}}
+\nu_{\tbox{2L}}^{\dag\tbox{(01)}}\sigma^{\mu}\partial_{\mu}\nu_{\tbox{2L}}^{\tbox{(01)}}
+\nu_{\tbox{2R}}^{\dag\tbox{(10)}}\overline{\sigma}^{\mu}\partial_{\mu}\nu_{\tbox{2R}}^{\tbox{(10)}}
+\nu_{\tbox{2R}}^{\dag\tbox{(01)}}\overline{\sigma}^{\mu}\partial_{\mu}\nu_{\tbox{2R}}^{\tbox{(01)}}\right)\nonumber\\
&\ \
-\frac{1}{R}\left(
(\nu_{\tbox{2L}}^{\dag\tbox{(10)}}\nu_{\tbox{2R}}^{\tbox{(10)}}
-\nu_{\tbox{2R}}^{\dag\tbox{(10)}}\nu_{\tbox{2L}}^{\tbox{(10)}})
-(\nu_{\tbox{2L}}^{\dag\tbox{(01)}}\nu_{\tbox{2R}}^{\tbox{(01)}}
-\nu_{\tbox{2R}}^{\dag\tbox{(01)}}\nu_{\tbox{2L}}^{\tbox{(01)}})\right).
\end{align}
%
%
>From this calculation, we see that we have two independent Dirac neutrinos that do
not mix with each other:
$\nu^{\tbox{(01)}}$ and
$\nu^{\tbox{(10)}}$.


Following the same procedure, we can find the normalization of
scalars.
\begin{align}
\Phi(+,+)&=\frac{1}{2\pi R}\phi^{(00)}+
\frac{1}{\sqrt{2}\pi R}\sum_{m,n}\cos\frac{m x^5+n x^6}{R}\phi^{(mn)},\nonumber\\
\Phi(+,-)&=\frac{1}{\sqrt{2}\pi R}\sum_{m,n}\cos\frac{m x^5+n x^6}{R}\phi^{(mn)},\nonumber\\
\Phi(-,+)&=\frac{1}{\sqrt{2}\pi R}\sum_{m,n}\sin\frac{m x^5+n x^6}{R}\phi^{(mn)},\nonumber\\
\Phi(-,-)&=\frac{1}{\sqrt{2}\pi R}\sum_{m,n}\cos\frac{m x^5+n
x^6}{R}\phi^{(mn)}.
\end{align}
Again, the rules of $m,n$ in the previous section for fermions apply
to the scalars.

\subsection{Gauge Bosons}
\newcommand{\mA}{\mathcal{A}}
As we are only interested in the normalization of the gauge fields, we
consider a generic gauge boson, $\mA_M$, associated with an $U(1)$ symmetry.
We then have the expansion
\begin{align}
\mathcal{L}_{gauge}&=-\frac{1}{4}F^{MN}F_{MN}\nonumber\\
&=-\frac{1}{4}F_{\mu\nu}F^{\mu\nu}-\frac{1}{2}F_{5\mu}F^{5\mu}
-\frac{1}{2}F_{6\mu}F^{6\mu}-\frac{1}{2}F_{56}F^{56}\nonumber\\
&=-\frac{1}{4}F_{\mu\nu}F^{\mu\nu}\nonumber\\
&\quad+\frac{1}{2}(\partial_{\mu}\mA_5\partial^{\mu}\mA_5
+\partial_{5}\mA_{\mu}\partial_{5}\mA^{\mu}
-\partial_{5}\mA_{\mu}\partial^{\mu}\mA_5
-\partial_{\mu}\mA_{5}\partial_{5}\mA^{\mu})\nonumber\\
&\quad+\frac{1}{2}(\partial_{\mu}\mA_6\partial^{\mu}\mA_6
+\partial_{6}\mA_{\mu}\partial_{6}\mA^{\mu}
-\partial_{6}\mA_{\mu}\partial^{\mu}\mA_6
-\partial_{\mu}\mA_{6}\partial_{6}\mA^{\mu})\nonumber\\
&\quad-\frac{1}{2}(\partial_5\mA_6\partial_5\mA_6
+\partial_6\mA_5\partial_6\mA_5 -2\partial_5\mA_6\partial_6\mA_5)
\label{eq:FKKExpand}
\end{align}
Notice that we have made the changes $\mA^{5,6}=-\mA_{5,6},
\partial^{5,6}=-\partial_{5,6}$, so that $\mA_{5,6}$ should be treated
as real, scalar fields.  We will work with this equation for the
various gauge particles.

As with the case of the neutral gauge bosons in our theory,
we assign the $Z_2\times Z_2^{\prime}$ parities to be
$\mA_{\mu}(++)$ and $\mA_{5,6}(--)$, and obtain the following lowest KK modes:
\begin{align}
A^{\tbox{(00)}}_{\mu},\ &A^{\tbox{(11)}}_{\mu},\ A^{\tbox{(20)}}_{\mu},\ A^{\tbox{(02)}}_{\mu},\nonumber\\
&A^{\tbox{(11)}}_{5},\ A^{\tbox{(20)}}_{5},\ A^{\tbox{(02)}}_{5},\nonumber\\
&A^{\tbox{(11)}}_{6},\ A^{\tbox{(20)}}_{6},\ A^{\tbox{(02)}}_{6},
\end{align}
and the KK-mode expansions for these states
\begin{align}
\mA_{\mu}&=
\frac{1}{2\pi R}
A^{\tbox{(00)}}_{\mu}+\frac{1}{\sqrt{2}\pi R}\left(
\cos\frac{2x^5}{R}A^{\tbox{(20)}}_{\mu}
+\cos\frac{2 x^6}{R}A^{\tbox{(02)}}_{\mu}
+\cos\frac{x^5+x^6}{R}A^{\tbox{(11)}}_{\mu}
\right),\nonumber\\
\mA_{5,6}&= \frac{1}{\sqrt{2}\pi R}\left(
\sin\frac{2x^5}{R}A^{\tbox{(20)}}_{5,6}
+\sin\frac{2 x^6}{R}A^{\tbox{(02)}}_{5,6}
+\sin\frac{x^5+x^6}{R}A^{\tbox{(11)}}_{5,6}\right).
\label{eq:B-LKKExpand}
\end{align}
One may check that the normalization gives canonical fields for
the scalars $A^{\tbox{KK}}_{5,6}$.

For the four-dimensional Lagrangian we insert the expansion of
Eq.~\ref{eq:B-LKKExpand} into Eq.~\ref{eq:FKKExpand} and integrate over
$x^5$ and $x^6$.  In additional to canonical kinetic energy terms,
we obtain the masses of these modes.
\begin{align}
\mathcal{L}&=\frac{1}{2}\left[
\frac{4}{R^2} A_{\mu}^{\tbox{(20)}} A^{\mu,\tbox{(20)}}+
\frac{4}{R^2} A_{\mu}^{\tbox{(02)}} A^{\mu,\tbox{(02)}}+
\frac{2}{R^2} A_{\mu}^{\tbox{(11)}} A^{\mu,\tbox{(11)}}
\right]\nonumber\\
&\!\! -\frac{1}{2}\left[
\frac{4}{R^2} (A^{\tbox{(20)}}_{5})^2+
\frac{4}{R^2} (A^{\tbox{(02)}}_{6})^2+
\frac{1}{R^2} (A^{\tbox{(11)}}_{5})^2+
\frac{1}{R^2} (A^{\tbox{(11)}}_{6})^2
- 2\frac{1}{R^2}A^{\tbox{(11)}}_{5}A^{\tbox{(11)}}_{6}\right]
\end{align}
We note first that we have some massless modes in $A_{6}^{\tbox{(02)}}$ and
$A_{5}^{\tbox{(20)}}$.  This can be traced to the fact that we do not have
terms $\partial_5 A_5$ and $\partial_6 A_6$ in $F_{\tbox{MN}}F^{\tbox{MN}}$.  As
in the case of 5D UED models, these modes are eaten by the
corresponding KK modes of $A_{\mu}^{\tbox{(mn)}}$ so they can become massive.
Here we note that the linear combination of
$\tfrac{1}{\sqrt{2}}(A^{\tbox{(11)}}_{5}+A^{\tbox{(11)}}_{6})$ is also massless, and is
eaten by $B_{\mu(11)}$.  Generally, at each KK level, one linear
combination of $A_{5}^{\tbox{KK}}$ and $A_{6}^{\tbox{KK}}$
(and any corresponding KK modes of Higgs particle, if
there is Higgs mechanism)
is eaten by $A_{\mu}^{\tbox{KK}}$,
while the orthogonal KK combination remains a physical mode, and is
a potential DM candidate if it is indeed the lightest KK mode.

\subsection{Normalization of Couplings}
In six-dimensional Lagrangian, both the yukawa and gauge couplings
are dimensionful.  We find the correct normalization by equating the
4D couplings to the effective 4D coupling resulting from integrating
over $x^5$ and $x^6$.  For example, consider a generic yukawa
interaction in the 6D theory
\begin{align*}
\mathcal{L}_{\tbox{6D-Yukawa}}=y^{\tbox{6D}}\overline{\Psi}_1\Phi \Psi_2
\end{align*}
where $y^{\tbox{6D}}$ has dimension $[M]^{-1}$.  The coupling
involving the zero-modes in the effective theory is then
\begin{align}
\mathcal{L}_{\tbox{4D-Yukawa}}
=
\int_{0}^{2\pi R}\!\!\!\!\!\!\!\!dx^5
\int_{0}^{2\pi R}\!\!\!\!\!\!\!\!dx^6
\frac{y^{\tbox{6D}}}{(2\pi R)^3}
\overline{\psi}^{\tbox{(00)}}_1\phi^{\tbox{(00)}}\psi_2^{\tbox{(00)}}
=
\frac{y^{\tbox{6D}}}{2\pi R}
\overline{\psi}^{\tbox{(00)}}_1\phi^{\tbox{(00)}}\psi_2^{\tbox{(00)}}.
\end{align}
So effectively we have $y^{\tbox{4D}}=y^{\tbox{6D}}(2\pi
R)^{-1}$.  Note that this is general: for the SM couplings in the 4D
effective theory, all fields are $(00)$ and have a normalization
$(2\pi R)^{-1}$, so the effective 4D couplings obtained after
integrating over $x^5$ and $x^6$ are simply the 6D couplings
multiplied by $(2\pi R)$.  By the same reasoning, we also have
$\lambda^{\tbox{6D}}=(2\pi R)^2 \lambda^{\tbox{4D}}$ for the
quartic coupling in the potential.

In general, the coupling between higher modes will come with extra
factors resulting from integrating over $x^5$ and $x^6$.  However,
the most important case for our purpose of calculating annihilation
diagrams involve couplings between two fermions and a boson where
exactly one of
field is a $(00)$ mode, and the two other fields are both $(mn)$
mode with $m,n$ nonzero.  Suppose we have a coupling in the 6D
Lagrangian of the form
$\mathcal{L}^{\tbox{6D}}=g^{\tbox{6D}}\overline{\Psi}\Phi\Psi$,
where $g^{\tbox{6D}}$ has dimension of $[M]^{-1}$, and we impose
that $g^{\tbox{6D}}=g^{\tbox{4D}}(2\pi R)$. In the 4D effective
theory we have
$\mathcal{L}^{\tbox{4D}}=g^{\tbox{4D}}\overline{\psi}^{\tbox{(00)}}\phi^{(mn)}\psi^{(mn)}$,
where the lower-case fields are the KK-modes of the corresponding 6D
fields in capital letters. The effective coupling between the KK modes
$g^{\tbox{4D}}\overline{\psi}^{\tbox{(00)}}\phi^{(mn)}\psi^{(mn)}$ in the
effective 4D theory is then
\begin{align}
\mathcal{L}_{\tbox{4D-eff}}=
\int_{0}^{2\pi R}\!\!\!\!\!\!\!\!dx^5
\int_{0}^{2\pi R}\!\!\!\!\!\!\!\!dx^6
\frac{g^{\tbox{4D}}(2\pi R)}
{(2\pi R)(\sqrt{2}\pi R)^2}c^2(m,n)
\overline{\psi}^{\tbox{(00)}}_1\phi^{(mn)}\psi^{(mn)}
=&g^{\tbox{4D}}\overline{\psi}^{\tbox{(00)}}_1\phi^{(mn)}\psi^{(mn)}.
\end{align}
So we see that there is no additional factors compared to the case
with all (00)-modes.


\end{document}